\documentclass[english,aps,prl,twocolumn,amsmath,amssymb,showpacs,superscriptaddress,notitlepage,longbibliography]{revtex4-1}
\usepackage[T1]{fontenc}
\usepackage[utf8]{inputenc}
\usepackage{mathrsfs}
\usepackage{graphicx}
\usepackage{esint}
\usepackage{xpatch}
\makeatletter

\providecommand{\tabularnewline}{\\}

\usepackage{amsmath}
\usepackage{amssymb}
\usepackage{bbm}
\usepackage{graphicx}
\usepackage[colorlinks=true,linkcolor=blue,anchorcolor=red,citecolor=blue,urlcolor=blue]{hyperref}
\usepackage[caption=false]{subfig}
\usepackage{lipsum}
\usepackage{balance}

\makeatother

\usepackage{babel}
\begin{document}
\title{Helical Topological Superconducting Pairing at Finite Excitation Energies}
\author{Masoud Bahari}
\email{masoud.bahari@physik.uni-wuerzburg.de}
\affiliation{Institute for Theoretical Physics and Astrophysics$\text{,}$ University
of W\"urzburg, D-97074 W\"urzburg, Germany}
\affiliation{W\"urzburg-Dresden Cluster of Excellence ct.qmat, Germany}
\author{Song-Bo Zhang}
\affiliation{Hefei National Laboratory, Hefei, Anhui, 230088, China}
\affiliation{International Center for Quantum Design of Functional Materials (ICQD), University of Science and Technology of China, Hefei, Anhui, 230026, China}
\author{Chang-An Li}
\affiliation{Institute for Theoretical Physics and Astrophysics$\text{,}$ University
of W\"urzburg, D-97074 W\"urzburg, Germany}
\affiliation{W\"urzburg-Dresden Cluster of Excellence ct.qmat, Germany}
\author{Sang-Jun Choi}
\affiliation{Department of Physics Education, Kongju National University, Gongju 32588, Republic of Korea}
\author{Philipp R\"u{\ss}mann}
\affiliation{Institute for Theoretical Physics and Astrophysics$\text{,}$ University
of W\"urzburg, D-97074 W\"urzburg, Germany}
\affiliation{W\"urzburg-Dresden Cluster of Excellence ct.qmat, Germany}
\affiliation{Peter Gr\"unberg Institut, Forschungszentrum J\"ulich and JARA, D-52425 J\"ulich, Germany}
\author{Carsten Timm}
\affiliation{W\"urzburg-Dresden Cluster of Excellence ct.qmat, Germany}
\affiliation{Institute of Theoretical Physics, Technische Universit\"at Dresden,
01062 Dresden, Germany}
\author{Bj\"orn Trauzettel}
\affiliation{Institute for Theoretical Physics and Astrophysics$\text{,}$ University
of W\"urzburg, D-97074 W\"urzburg, Germany}
\affiliation{W\"urzburg-Dresden Cluster of Excellence ct.qmat, Germany}
\date{\today}
\begin{abstract}
We propose helical topological superconductivity away from the Fermi surface in three-dimensional time-reversal-symmetric
odd-parity multiband superconductors. In these systems, pairing between electrons originating from different
bands is responsible for the corresponding topological phase transition.
Consequently, a pair of helical topological Dirac
surface states emerges at finite excitation energies. These helical Dirac
surface states are tunable in energy by chemical potential and strength of band-splitting. They are protected by time-reversal symmetry combined with crystalline two-fold rotation symmetry. We suggest concrete materials in which this phenomenon
could be observed.
\end{abstract}
\maketitle
\textit{Introduction.}\textemdash 
The search for intrinsic topological superconductors has been an active research area in the condensed matter community for many years \citep{Review-2013,Review-2015-1,Review-2015-2,Review-2017,Review-2018}. In weakly interacting systems, the essential ingredient for intrinsic topological phase transitions (TPTs) is a weak attractive electron-electron interaction appearing close to the Fermi surface. This results in particular bulk properties including topological superconducting gaps \citep{Gapped-2008,Gapped-2009,Gapped-2009-1,Gapped-2010-1,Gapped-2010-2,Gapped-2010-3,Gapped-2011,PairingMatrices-2015,PairingMatrices-2016,HOTI-super-2018-1,HOTI-super-2018-2,HOTI-super-2019,HOTI-super-2020},
possibly with nodes \citep{NodeLine-2002,NodeLine-2006,NodeLine-2010,NodeLine-2014,NodeLine-2015,2DNodes,CongjunPRB,PairingMatrices-2018,PairingMatrices-2020,NodeLine-2021}. Due to particle-hole symmetry, the bulk-boundary
correspondence allows for emergent Majorana boundary states protected by band topology. 

Finite energy (FE) Cooper pairing is a particular type of pairing
that can happen away from the Fermi surface \citep{FinieEnergyPairing-2009,FinieEnergyPairing-2015,FinieEnergyPairing-2017,FEPairingSymmetry-2019,FinieEnergyPairing-2021,
FinieEnergyPairing-2022-1,FinieEnergyPairing-2022-2,FinieEnergyPairing-2022-3,FinieEnergyPairing-2022-1,Masoud2023}. It arises from multiband
effects in superconductors  \citep{Multiband-1980,Multiband-2011,Multiband-2014-1,Multiband-2014-2,Multiband-2016-1,Multiband-2016-2,Multiband-2017,Multiband-2018-1,
Multiband-2018-2,Multiband-2019-1,Multiband-2019-2,Multiband-2019-3,Multiband-2019-4,Multiband-2020,Multiband-2021-1,Multiband-2021-2,Multiband-2021-3,
Multiband-2022}. Dips in the density of states at FEs is a particular spectral feature of such pairing found in hybrid structures based on the superconducting
proximity effect of $\text{NbSe}_{2}$ in the topological insulators
$\text{\text{BiSbTe}}_{1.25}\text{Se}_{1.75}$ \citep{GLS3} and $\text{Bi}_{2}\text{Se}_{3}$
\citep{GLS2}. In this Letter, we show that TPTs can be exclusively induced by FE Cooper pairing emerging from
time-reversal-symmetric odd-parity multiband superconductors. The interplay of spin-orbit coupling and electron-phonon interaction allows for the formation of unconventional FE Cooper pairing with topologically nontrivial order. Bulk band
crossings are partially gapped out in momentum space at FEs due to odd-parity superconductivity. This leads to TPTs as evidenced by
helical topological Dirac surface states at finite excitation energies. Such surface states are composed of electrons and holes with different magnetic quantum numbers. They are distinct from conventional Majorana boundary states due to broken particle-hole symmetry away from the Fermi energy. The helical Dirac points are (i) protected topologically
by time-reversal $\hat{T}$ symmetry combined with crystalline two-fold rotation as we specify below, and (ii) tunable in energy by chemical potential and strength
of band splitting.

To observe FE Cooper pairing, the normal state should have at least
two energy bands with the same sign of curvature close to the Fermi
energy. The bands need to have different effective masses. Therefore,
our results are relevant for a broad range of multiband superconductors as we elaborate on below.

To capture the underlying physics, we develop a theory for generic
superconducting multiband systems with four-valued local degrees of
freedom. For concreteness, we work with a model Hamiltonian describing
the band structure of electrons with effective angular momentum $j=3/2$
near the Fermi energy. However, the formalism is not restricted to this particular choice of angular momentum quantum numbers.

\textit{Model.}\textemdash We start with the Luttinger-Kohn Hamiltonian
$\hat{\mathcal{H}}_{{\bf k}}=\alpha|{\bf k}|^{2}+\beta({\bf k}\cdot\mathbf{J})^{2}-\mu$,
which describes $j=3/2$ electrons within the $\Gamma_{8}$ bands \citep{2DNodes,Brydon,Luttinger1,Luttinger2}.
The kinetic, symmetric spin-orbit-coupling, and chemical  potential terms are parameterized by $\alpha$, $\beta$, and $\mu$, respectively.
${\bf k}=(k_{x},k_{y},k_{z})$ is the three-dimensional (3D) momentum and $\mathbf{J}=(\hat{J}_{x},\hat{J}_{y},\hat{J}_{z})$
are the $4\times4$ spin matrices in the $j=3/2$ formalism. The normal
state has a pair of doubly degenerate quadratic energy bands given
by $E_{{\bf k}}^{+}=(\alpha+9\beta/4)|{\bf k}|^{2}-\mu$ and $E_{{\bf k}}^{-}=(\alpha+\beta/4)|{\bf k}|^{2}-\mu$. 
The chemical potential $\mu$ should fulfill $|\mu|<N\hbar\omega_{D}$, where $\omega_{D}$ is the Debye frequency and $N=1.25+\alpha/\beta$ is a material dependent number. While $\hat{\mathcal{H}}_{{\bf k}}$ preserves O(3) symmetry, our predictions are generic and can be applied to systems with relevant point group symmetry in the normal state \citep{Comment_Othree}.

\begin{figure}[t]
\begin{centering}
\includegraphics[scale=0.85]{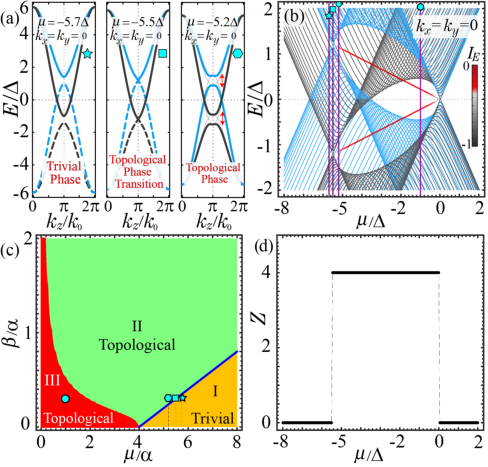}
\par\end{centering}
\caption{\label{Fig1} (a) BdG energy spectra along the {[}001{]} direction
for three different values of the chemical potential. Solid (dashed) lines depict twofold degenerate particle (hole) bands. Decoupled energy bands are plotted with distinct colors (blue and black) due to $\hat{C}_{P}$ symmetry. In the third panel, the fade gray lines denote $\Delta=0$. (b) Spectrum versus $\mu$ for a (001) slab with $80$ layers. The color bar indicates
the inverse participation ratio \citep{Comment_IPR}. The gray and blue colors represent bulk states, and red color denotes edge states. (c) Topological phase diagram induced by FE pairing. Region
III (II) indicates that the helical Dirac points are located (coexist with) within
the low energy gap (bulk states). (d) Topological invariant corresponding
to panel (b). $A_{2u}$ pairing is responsible for superconductivity
in all panels. Other parameters are $\beta=0.3\alpha$, $\alpha=-\Delta$
and $k_{0}=a^{-1}$ with $a$ being the lattice constant in corresponding tight-binding
calculations.}
\end{figure}
The superconducting properties of the system can be captured by the
Bogoliubov-de Gennes (BdG) Hamiltonian
\begin{equation}
\hat{H}({\bf k})=\left(\begin{array}{cc}
\hat{\mathcal{H}}_{{\bf k}} & \hat{\Delta}_{{\bf k}}\\
\hat{\Delta}_{{\bf k}}^{\dagger} & -\hat{\mathcal{H}}_{-{\bf k}}^{T}
\end{array}\right),\label{BdG Ham}
\end{equation}
where $\hat{\mathcal{H}}_{{\bf k}}$ is the inversion-symmetric normal
part and $\hat{\Delta}_{{\bf k}}$ is a $4\times4$ pairing matrix
describing pairing of electrons with four-valued degrees of freedom.
Generally, the explicit form of $\hat{\Delta}_{{\bf k}}$ depends
on the basis of the system. For concreteness, we focus on systems
with cubic point group $O_{h}$. In this case, $\hat{\Delta}_{{\bf k}}$
can be expanded into irreducible tensor operators of irreducible representations (irreps) of $O_{h}$
\citep{LucilePRB,VenderbosPRX,Jiabin,FinieEnergyPairing-2022-1}.

\textit{Topology at finite energies.}\textemdash The prerequisite
for TPTs away from the Fermi energy is that the $E_{{\bf k}}^{+}$
electron and $-E_{{\bf k}}^{-}$ hole bands cross each other. These
band crossings become anti-crossings in the presence of odd-parity
FE Cooper pairing as we explain in detail in the Supplemental Material (SM) \citep{Supplemental_Material}.  For concreteness, we consider
the $A_{2u}$ ($p$-wave septet) pairing given by $\hat{\Delta}_{{\bf k}}=\Delta(\mathbf{k}\cdot\hat{\mathbf{T}})\hat{\mathcal{R}}$,
where $\Delta$ is the pairing strength, $\hat{\mathcal{R}}=e^{i\pi\hat{J}_{y}}$
is the fermionic antisymmetry factor, and $\hat{T}_{i}=\{\hat{J}_{i},\hat{J}_{i+1}^{2}-\hat{J}_{i+2}^{2}\}$
with $i+1=y$ if $i=x$, etc., cyclically \citep{Brydon,LucilePRB}.
Note that $p$-wave septet pairing has been argued to be the favorable
pairing symmetry in systems with cubic symmetry \citep{Brydon,Exp_Hyunsoo,TuningParityExp}.
Although we focus on this specific pairing channel, our qualitative
results are general. They apply to several odd-parity pairing
channels summarized in Table. \ref{Table1}.
\begin{table}[t]
\begin{tabular}{|c|c|c|c|}
\hline 
$O_{h}$ &  & ${\bf k}$ & $\hat{\Delta}_{{\bf k}}^{+-}=\Delta\times$\tabularnewline
\hline 
$E_{u}$ & $E_{u}^{(2)}$ & $(0,0,k_{z})$ & $k_{z}\hat{\tau}_{0}$\tabularnewline
$T_{2u}$ & $T_{2u}^{(1)}$ & $(0,0,k_{z})$ & $k_{z}\hat{\tau}_{z}$\tabularnewline
 & $T_{2u}^{(2)}$ & $(0,k_{y},0)$ & $k_{y}\hat{\tau}_{x}$\tabularnewline
 & $T_{2u}^{(3)}$ & $(k_{x},0,0)$ & $ik_{x}\hat{\tau}_{x}$\tabularnewline
\hline 
$A_{2u}$ &  & $(k_{x},0,0)$ & $-k_{x}\hat{\tau}_{0}$\tabularnewline
 &  & $(0,k_{y},0)$ & $-ik_{y}\hat{\tau}_{z}$\tabularnewline
 &  & $(0,0,k_{z})$ & $k_{z}\hat{\tau}_{0}$\tabularnewline
$T_{2u}$ & $T_{2u}^{(1)}$ & $(0,0,k_{z})$ & $k_{z}\hat{\tau}_{z}$\tabularnewline
 &  & $(\pm k,\pm k,0)$ & $k(\pm\hat{\tau}_{0}\pm i\hat{\tau}_{z})$\tabularnewline
 & $T_{2u}^{(2)}$ & $(k_{x},0,0)$ & $k_{x}\hat{\tau}_{x}$\tabularnewline
 &  & $(0,\pm k,\pm k)$ & $k(\hat{\tau}_{y}\pm\hat{\tau}_{z}/3)$\tabularnewline
 & $T_{2u}^{(3)}$ & $(0,k_{y},0)$ & $k_{y}\hat{\tau}_{x}$\tabularnewline
 &  & $(\pm k,0,\pm k)$ & $k(\pm\hat{\tau}_{0}/3-i\hat{\tau}_{y})$\tabularnewline
\hline 
\end{tabular}
\caption{\label{Table1}TPTs at FEs in $p$-wave pairing channels distinguished
by cubic irreps (first column) and their components (second column).
The direction of TPTs is denoted in the third column. The last column
indicates the FE pairing potential responsible for TPTs. The pairing
matrices for the given irreps are given in the SM \citep{Supplemental_Material,Comment_Lucile}.}
\end{table} 

Without loss of generality, we illustrate our predictions for the {[}001{]}
direction (i.e., $k_{x}=k_{y}=0$). We observe in Fig.~\ref{Fig1}(a)
that the bulk bands, i.e., electron $E_{k_{z}}^{+}$ and hole $-E_{k_{z}}^{-}$
branches, do not touch each other for $\mu=-5.7\Delta$ close to $k_{z}=\pi$
away from the Fermi surface. In this case, the system exhibits a topologically
trivial phase. Increasing slightly the Fermi energy to $\mu=-5.5\Delta$,
this makes the two bands touch each other. For $\mu=-5.2\Delta$,
the presence of FE pairing, i.e., pairing between electrons with different
magnetic quantum numbers ($|m_{j}|=3/2$ with $|m_{j}|=1/2$), opens
a band gap indicating a TPT. This TPT is accompanied by the emergence
of helical topological surface states. To illustrate the appearance of helical surface
states (HSSs), the spectrum of a finite system is depicted in Fig.~\ref{Fig1}(b) \citep{Comment_Tightbinding}. The localized fourfold degenerate HSSs are marked by red
color indicating large inverse participation ratio (IPR) \citep{Comment_IPR}. They emerge within the range of Fermi energies $\mu/\Delta\in[-5.5,0]$,
in which a nonzero topological invariant, see Fig.~\ref{Fig1}(d),
can be found. We describe below how the topological invariant is defined. Generally, when the energy bands have downward (upward)
curvature, i.e., $\text{sgn}(\alpha)=\text{sgn}(\beta)=-(+)1$, the
topological phase occurs in the range $\mu/\Delta\in[\mu_{c},0]\ ([0,\mu_{c}])$
with the critical value $\mu_{c}=4\alpha+5\beta$. The analysis of the spectrum along other directions reveals that these HSSs are helical topological Dirac points occurring at finite excitation energies as shown in Fig.~\ref{Fig2}. Notably, the helical surface points do not hybridize with the bulk states due to the combination of time-reversal and two-fold rotation symmetries \citep{Comment_BIC}.

According to the topological phase diagram, depicted in the
plane of $(\mu/\alpha,\beta/\alpha)$ in Fig.~\ref{Fig1}(c),
the FE Cooper pairing in odd-parity multiband superconductors is intrinsically
topological (regions II and III) within a certain range of Fermi energies.

\textit{Theory for topological phase transition.}\textemdash The nontrivial
topology at FEs originates from unconventional pairing of FE electrons.
To understand the underlying mechanism, we project the BdG Hamiltonian
onto the doubly degenerate eigenspinors of the normal state $\hat{V}_{{\bf k}}^{\pm}$
as the pseudospin basis. In this basis, the inter-band representation
of the BdG Hamiltonian becomes
\begin{equation}
\hat{\mathscr{H}}({\bf k})=\Bigg(\begin{array}{cc}
\hat{\mathsf{H}}_{{\bf k}}^{+-} & \hat{\Delta}_{{\bf k}}^{\text{intra}}\\
\big(\hat{\Delta}_{{\bf k}}^{\text{intra}}\big)^{\dagger} & \hat{\mathsf{H}}_{{\bf k}}^{-+}
\end{array}\Bigg),\label{InterBandH}
\end{equation}
where $\hat{\mathsf{H}}_{{\bf k}}^{+-}$ and $\hat{\mathsf{H}}_{{\bf k}}^{-+}$ denote the interband parts of the BdG Hamiltonian and $\hat{\Delta}_{{\bf k}}^{\text{intra}}$ describes intraband pairing
at the Fermi energy. The subblock matrices in Eq. $\eqref{InterBandH}$
can be written as
\begin{equation}
\!\!\!\hat{\mathsf{H}}_{{\bf k}}^{+-}=\left(\!\!\!\begin{array}{cc}
E_{{\bf k}}^{+} & \hat{\Delta}_{{\bf k}}^{+-}\\
(\hat{\Delta}_{{\bf k}}^{+-})^{\dagger} & -E_{{\bf k}}^{-}
\end{array}\!\!\!\right),\ \ \hat{\Delta}_{{\bf k}}^{\text{intra}}\!=\!\left(\!\!\!\begin{array}{cc}
0 & \hat{\Delta}_{{\bf k}}^{++}\\
(\hat{\Delta}_{{\bf k}}^{--})^{\dagger} & 0
\end{array}\!\!\!\right),\!\!\label{hplusminus}
\end{equation}
where $\hat{\Delta}_{{\bf k}}^{\nu\nu^{\prime}}$ with $\nu,\nu^{\prime}\in\{+,-\}$  denotes the pairing
potential projected onto the normal state bands defined by $\hat{\Delta}_{{\bf k}}^{\nu\nu^{\prime}}=\hat{V}_{{\bf k}}^{\nu\dagger}\hat{\Delta}_{{\bf k}}(\hat{V}_{-{\bf k}}^{\nu^{\prime}\dagger})^{T}$
where $\nu\neq\nu^{\prime}$ ($\nu=\nu^{\prime}$) indicates inter-
(intra-) band pairing between electrons having different
(identical) magnetic quantum numbers in magnitude.
\begin{figure}[t]
\begin{centering}%
\includegraphics[scale=0.9]{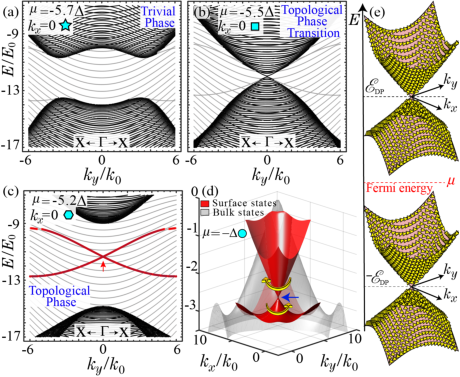}
\par\end{centering}
\caption{\label{Fig2} Spectra of a slab with (001) surfaces as a functions of $k_{y}$ indicating (a) topologically trivial phase, (b)   topological phase transition, and (c) topological phase at FEs. The thickness is 200 layers. (d)
3D energy dispersion versus $k_{x}$ and $k_{y}$. (e) Helical surface cones at positive and negative energies related by particle-hole symmetry. (d,e) Yellow arrows indicate left- (right-) handed spin texture in the upper (lower) cone. $E_{0}=0.1\Delta$ and other parameters
are the same as in Fig.~\ref{Fig1}.}
\end{figure}
In our model, the most convenient way to explain TPTs at FEs is to make
$\hat{\mathscr{H}}({\bf k})$ block diagonal. Thus, TPTs occur in directions in momentum space where the odd-parity interband pairing is finite,
whereas the intraband pairing vanishes (nodes at the Fermi energy), i.e., $\hat{\Delta}_{{\bf k}}^{\nu\nu^{\prime}}\neq0$
and $\hat{\Delta}_{{\bf k}}^{\text{intra}}=0$. These conditions are satisfied, particularly for the $A_{2u}$ pairing, when time-reversal symmetry combines with a two-fold rotation around the $\langle110\rangle$ direction. This combination is expressed as $\hat{C}_{T}=\hat{T}\hat{C}_{2,x+y}$ yielding $\hat{C}_{T}^{2}=1$ \citep{VenderbosPRX,Supplemental_Material,CommentsCT,CT_2015,CT_2018,CT_2019,CT_2020,CT_2021_1,CT_2021_2}. Consequently, the point nodes emerge at the Fermi energy accompanied by nonvanishing interband pairing along the $\langle001\rangle$ direction. In this case, Eq.
$\eqref{InterBandH}$ becomes block diagonal and $\hat{\mathsf{H}}_{{\bf k}}^{\nu\nu^{\prime}}$ is further reducible. To block diagonalize $\hat{\mathsf{H}}_{{\bf k}}^{\nu\nu^{\prime}}$,
we define a general rotation symmetry operator in interband basis
\begin{equation}
\mathcal{\hat{D}}_{\mathbf{n}}(\theta,\epsilon)\equiv\text{diag}\big(e^{i\theta/2(\mathbf{n}\cdot\hat{\boldsymbol{\tau}})},\epsilon e^{-i\theta/2(\mathbf{n}\cdot\hat{\boldsymbol{\tau}})^{*}}),\label{SpinPi_rot_BdG}
\end{equation}
where $\theta$ is the rotation angle about an arbitrary unit
vector $\mathbf{n}$. The vector of Pauli matrices
in pseudospin basis is $\hat{\boldsymbol{\tau}}=(\hat{\tau}_{x},\hat{\tau}_{y},\hat{\tau}_{z})$
and $\epsilon=\pm1$ stands for two unitarily equivalent representations. Along the TPT directions, the superconducting Hamiltonian of each block preserves a two-fold rotation by $\theta=\pi$, i.e., $[\mathcal{\hat{D}}_{\mathbf{n}}(\pi,\epsilon),\hat{\mathsf{H}}_{{\bf k}}^{\nu\nu^{\prime}}]=0$.
This allows us to label the corresponding energy bands by the sign $\pm$ corresponding to the eigenvalues $\lambda=\pm i$
of $\mathcal{\hat{D}}_{\mathbf{n}}(\pi,\epsilon)$.
This operation satisfies the property $\mathcal{\hat{D}}_{\mathbf{n}}^{2}(\pi,\epsilon)=-\hat{\mathbbm{1}}$ \cite{Comment_twofold}. $\hat{\mathsf{H}}_{{\bf k}}^{\nu\nu^{\prime}}$ can be reduced to two independent subblock matrices in the eigenspaces of $\mathcal{\hat{D}}_{\mathbf{n}}(\pi,\epsilon)$.
The corresponding similarity transformation is $\hat{\mathcal{U}}^{\dagger}\hat{\mathsf{H}}_{{\bf k}}^{\nu\nu^{\prime}}\hat{\mathcal{U}}=\text{diag}\big(\hat{h}_{{\bf k},-}^{\nu\nu^{\prime}},\hat{h}_{{\bf k},+}^{\nu\nu^{\prime}}\big)$,
where $\hat{\mathcal{U}}$ is the unitary matrix defined by the eigenvectors
of $\mathcal{\hat{D}}_{\mathbf{n}}(\pi,\epsilon)$. Then, $\hat{h}_{{\bf k},\lambda}^{\nu\nu^{\prime}}$
becomes
\begin{align}
\hat{h}_{{\bf k},\lambda}^{\nu\nu^{\prime}} & =\left(\begin{array}{cc}
-E_{{\bf k}}^{\nu^{\prime}} & \big(\delta_{{\bf k},\lambda}^{\nu\nu^{\prime}}\big)^{*}\\
\delta_{{\bf k},\lambda}^{\nu\nu^{\prime}} & E_{{\bf k}}^{\nu}
\end{array}\right),\label{FinalDecopled}
\end{align}
where $\delta_{{\bf k},\lambda}^{\nu\nu^{\prime}}$ is the FE pairing
represented in the eigenbasis of $\mathcal{\hat{D}}_{\mathbf{n}}(\pi,\epsilon)$ \citep{Comment_Hamil}. Eventually, the transformed BdG Hamiltonian, Eq. $\text{\eqref{InterBandH}}$, decouples into four $2\times2$ subblocks, 
\begin{equation}
\hat{\mathscr{H}}^{\prime}({\bf k})=\text{diag}\big(\hat{h}_{{\bf k},+}^{+-},\hat{h}_{{\bf k},-}^{+-},\hat{h}_{{\bf k},+}^{-+},\hat{h}_{{\bf k},-}^{-+}\big).\label{Blockdiag}
\end{equation}
Note that $\hat{\mathscr{H}}^{\prime}({\bf k})$ illustrates the decoupling of the double degeneracy of states. For the $A_{2u}$ pairing channel, this is achieved through the mirror reflection symmetry $\hat{C}_{P}=\hat{P}\hat{C}_{2,x+y}$, where $\hat{P}$ is inversion symmetry, yielding $\hat{C}_{P}^{2}=-1$ (see Fig. \ref{Fig1}(a)).

The topological properties at FEs are encoded in the subblocks of
$\hat{\mathscr{H}}^{\prime}({\bf k})$. Generically, when the pairing
term in $\hat{h}_{{\bf k},\lambda}^{\nu\nu^{\prime}}$ is odd in momentum,
the FE bulk band gaps between electron and hole branches cross
or anticross at the parity-invariant momenta. This happens through
the tuning of the normal state parameters, i.e., $(\alpha,\beta,\mu)$,
and leads to a topologically nontrivial phase as described below.

The topological classification of our 3D model in class DIII is captured by a $\mathbb{Z}$ topological index at low energies \citep{Comment_TopolClass,Topological-Classification-2007,Topological-Classification-2008,Topological-Classification-2011,Topological-Classification-2013,Topological-Classification-2016}. We are, however, interested in a different type of topological classification at finite excitation energies. This classification is based on the sum of four $\mathbb{Z}_2$ indices stemming form a quantization of the Berry phase defined for each block presented in Eq. \eqref{Blockdiag}. Note that $\hat{h}_{{\bf k},-}^{+-}$ is essentially a replica of  $\hat{h}_{{\bf k},+}^{+-}$,  owing to the presence of  $\mathcal{\hat{D}}_{\mathbf{n}}(\pi,\epsilon)$ symmetry. Additionally, $\hat{h}_{{\bf k},\pm}^{-+}$ represent the particle-hole counterparts of $\hat{h}_{{\bf k},\pm}^{+-}$. Therefore, the decoupled sectors in Eq. \eqref{Blockdiag} are topologically identical. The topological invariant is given by $Z=\sum_{\lambda\nu\nu^{\prime}}\ensuremath{N_{\lambda}^{\nu\nu^{\prime}}}$ with $\mathbb{Z}_2$ index $N_{\lambda}^{\nu\nu^{\prime}}=(1/\pi)\oint_{c}d{\bf k}\cdot i\langle u_{{\bf k},\lambda}^{\nu\nu^{\prime}}|\boldsymbol{\nabla}_{{\bf k}}|u_{{\bf k},\lambda}^{\nu\nu^{\prime}}\rangle$, where $|u_{{\bf k},\lambda}^{\nu\nu^{\prime}}\rangle$ is the eigenspinor
of $\hat{h}_{{\bf k},\lambda}^{\nu\nu^{\prime}}$ associated with
the lower energy band \citep{Berry-1,Berry-2,Berry-3,Berry-4}. In
the topological phase, $N_{\lambda}^{\nu\nu^{\prime}}$ is quantized due to parity symmetry implied by $\hat{\tau}_z\hat{h}_{{\bf k},\lambda}^{\nu\nu^{\prime}}\hat{\tau}^{-1}_z=\hat{h}_{{-\bf k},\lambda}^{\nu\nu^{\prime}}$. We obtain analytically that $\hat{h}_{{\bf k},\lambda}^{\nu\nu^{\prime}}$ is topologically nontrivial with $N_{\lambda}^{\nu\nu^{\prime}}=1$, when $\mu/\Delta\in[\mu_{c},0]\ ([0,\mu_{c}])$ for $\text{sgn}(\alpha)=\text{sgn}(\beta)=-(+)1$.
The topological phase diagram in Fig.~\ref{Fig1}(c) illustrates
this analysis \cite{Comment_Table}. The topological index
becomes a net value of $Z=4$ indicating the emergence of two pairs of helical Dirac states on the surface of superconductor, cf. Fig. \ref{Fig2}(e). One helical pair appears at positive excitation energy while the other pair emerges at negative excitation energy \citep{Comment_FourfoldDeg}. 

\textit{2D surface states.}\textemdash{} The HSSs disperse linearly in the vicinity of the $\Gamma$ point of the 2D surface
Brillouin zone as illustrated in Fig.~\ref{Fig2}(c) and
\ref{Fig2}(d). At larger momenta, the surface dispersion connects to the bulk states. This is due to the intraband pairing at the Fermi energy, i.e., $\hat{\Delta}_{{\bf k}}^{\text{intra}}$. It is finite for nonvanishing $k_{x}$ or $k_{y}$. Hence, $\hat{\mathscr{H}}^{\prime}({\bf k})$
is no longer block-diagonal. This lifts the twofold degeneracy except at the Dirac
points, which is the reason for the emergence of helical Dirac surface cones at FEs.

To gain insight into the dispersion of the HSSs, we consider
an open surface perpendicular to the $z$ direction and derive an effective
Hamiltonian for the HSSs near the Dirac points. We consider
$k_{x}$ and $k_{y}$ small and decompose the BdG Hamiltonian as $\hat{H}({\bf k})=\hat{H}(0,0,k_{z})+\delta\hat{H}(k_{x},k_{y},0)$.
First, we derive the Dirac point energy at $k_{x}=k_{y}=0$. At this
point, $\hat{H}(0,0,k_{z})$ becomes block diagonal. This
leads to $\hat{\mathscr{H}}^{\prime}(0,0,k_{z})=\text{diag}(\hat{h}_{k_{z},+}^{+-},\hat{h}_{k_{z},-}^{+-},\hat{h}_{k_{z},+}^{-+},\hat{h}_{k_{z},-}^{-+})$, 
which is identical to Eq. $\text{\eqref{Blockdiag}}$. We consider
a semi-infinite system in the half space $z\geq0$ with open boundary
conditions. Therefore, $k_{z}$ is no longer conserved and we treat
the $z$ direction in real space, where the momentum operator is $-i\partial_{z}$.
The eigenvalue equation for each subblock reads $\hat{h}_{-i\partial_{z},\lambda}^{\nu\nu^{\prime}}\hat{\Phi}(\xi,z)=\mathscr{E}_{\text{DP}}\hat{\Phi}(\xi,z)$,
where $\nu,\nu^{\prime}\in\{+,-\}$ and $\hat{\Phi}(\xi,z)=(u,v)^{T}\text{exp}(\xi z)$
is the trial eigenspinor of the Dirac point with $\xi$ being its
penetration length. $|u|^{2}$ ($|v|^{2}$) are probability weights
for electron (hole) bands having different band indices. We derive the energy of the helical topological Dirac surface points induced by unconventional FE pairing as
\begin{equation}
\mathscr{E}_{\text{DP}}=\pm\mu\frac{m-m^{\prime}}{m+m^{\prime}},
\end{equation}
 where $m=\alpha+\beta/4$ and $m^{\prime}=\alpha+9\beta/4$ are the
masses for the normal state bands. Notably, the Dirac points are tunable
in energy by chemical potential $\mu$ and spin-orbit-coupling strength
$m-m^{\prime}=-2\beta$ \citep{com_generic}. 

Projecting the BdG Hamiltonian onto the basis of surface states results in an effective Hamiltonian for the FE helical topological
surface states of the form \citep{Supplemental_Material}
\begin{align}
\hat{\mathtt{H}}(k_{x},k_{y}) & =(\mathscr{E}_{\text{DP}}-\varsigma_{1}{\bf k}_{||}^{2})\hat{\sigma}_{0}+\varsigma_{2}(k_{y}\hat{\sigma}_{x}-k_{x}\hat{\sigma}_{y}),\label{Dirac cone}
\end{align}
where ${\bf k}_{||}^{2}=k_{x}^{2}+k_{y}^{2}$, and $\varsigma_{1(2)}$
is the group velocity of the HSSs. To leading order, the HSSs exhibit the Dirac dispersion $\mathcal{E}_{\pm}\approx\mathscr{E}_{\text{DP}}\pm\varsigma_{2}|\mathbf{k}_{||}|$. In the limit, where either $\mu$ or $m-m^{\prime}$ vanishes, the Dirac points shift to zero energy. Then, the FE helical topological surface states become dispersive helical Majorana modes \citep{Review-2013,Review-2015-1,Review-2015-2,Review-2017,Review-2018}.
In this case, $\hat{h}_{k_{z},\pm}^{+-}$ ($\hat{h}_{k_{z},\pm}^{-+}$) resembles
an ordinary $p$-wave superconducting Hamiltonian \citep{Kitaev}. However, for finite $\mu$ and $m-m^{\prime}$, topological HSSs emerge at FE. 

To identify a proper material to observe this phenomenon, certain conditions need to be satisfied.
In the normal state: (i) relevant energy bands should have the same
signs of curvature close to the Fermi energy, i.e., both curving downward or upward, (ii) time-reversal symmetry and twofold rotation should both be present. In the superconducting state: (i)  pairing potential should be of odd-parity type, (ii) nodes should be present in the BdG spectrum at the Fermi energy such as in $\text{UTe}_{2}$ \citep{PointN2}, (iii) nonvanishing pairing at FEs should be allowed. Candidates are antiperovskite oxides $\text{Sr}_{3-x}\text{SnO}$ \citep{Ex1}  
and half-Heusler materials \textit{R}PdBi with \textit{R$\,\in$}\{Y,Dy,Tb,Sm\}
\citep{Ex2,Comment_ASOC}. 
Specifically, weakly hole-doped YPdBi is a promising material to observe unconventional FE Cooper pairing \citep{Ex2,Radmanesh_1}. In the SM \citep{Supplemental_Material}, we combine density functional theory \citep{Philipp_1,Philipp_2,Philipp_3_1,Philipp_4} and analytical model analysis to estimate the magnitude of FE pairing in holed-doped YPdBi \citep{Supplemental_Material,Masoud2023,DFT_ScientificReports}. We obtain an energy range of $\Delta_{E}\approx7.7-46.2$ \textit{\textmu}eV. Notably, an energy resolution below 8 \textit{\textmu}eV at operating temperatures of 10 mK is possible via state-of-the-art STM and transport experiments in dilution refrigerators \citep{LowTemp}. Hence, FE Cooper pairing should be observable in hole-doped YPdBi.

In addition to the materials mentioned above, hybrid structures
such as $\text{\text{BiSbTe}}_{1.25}\text{Se}_{1.75}/\text{NbSe}_{2}$
\citep{GLS3,NbSe2}, and $\text{Bi}_{2}\text{Te}_{3}/\text{NbSe}_{2}$
\citep{Bi2Te3DFT,NbSe2}, and \textit{X}/\textit{S}Bi with \textit{S$\,\in$}\{YPt,LuPd\} \citep{Brydon,Exp_Hyunsoo,TuningParityExp},
and \textit{X$\,\in$}\{Si,Ge\} are suitable candidates for FE pairing. The excitations
in Si, Ge \citep{Book_semiconductor_2006}, and \textit{S}Bi
\citep{Brydon,TuningParityExp} have $j=3/2$ character near the Fermi
energy with suitable curvature.
The pairing order parameter in \textit{S}Bi is believed to have $p$-wave
septet symmetry \citep{Exp_Hyunsoo,TuningParityExp}. Therefore, the
superconducting proximity effect of\textit{ S}Bi to Si or Ge should induce FE pairing signaled by the emergence of HSSs. 

\textit{Conclusions.}\textemdash{} We have shown that helical topological
superconducting pairing emerges in multiband time-reversal-symmetric odd-parity
superconductors due to unconventional pairing away from the Fermi energy. This leads to the appearance of tunable helical
topological Dirac surface states at FEs. They are topologically protected against perturbations due to combination of time-reversal and two-fold rotation symmetries \cite{Supplemental_Material}. 
Promising experimental probes are (spin-polarized) angle-resolved photoemission 
\citep{ARPES-2009,ARPES-2011,ARPES-2014,ARPES-2019,ARPES-2020-1,ARPES-2020-2,ARPES-2021-1,ARPES-2021-2,ARPES-2022}
and scanning tunneling spectroscopies \citep{STM-2003,STM-2004,STM-2010,ARPES-STM-2007,STM-2014,STM-2016,STM-2017-1,STM-2017-2,STM-2021-1,STM-2021-2,STM-2021-3}. 
\begin{acknowledgements}
The work was supported by the DFG (SPP 1666, SFB 1170 ToCoTronics, and SFB 1143, project A04, Project-Id 247310070), and the W\"urzburg-Dresden Cluster of Excellence ct.qmat,
EXC 2147, Project-Id 390858490. We thank the Bavarian Ministry of
Economic Affairs, Regional Development and Energy for financial support
within the High-Tech Agenda Project \textquotedblleft Bausteine f\"ur
das Quanten Computing auf Basis topologischer Materialen\textquotedblright.

\end{acknowledgements}

\providecommand{\noopsort}[1]{}\providecommand{\singleletter}[1]{#1}%
%

  \appendix
\begin{center}
\bf{\large Supplemental Material}
\end{center}
\maketitle
In this section, we address why TPTs at FEs occur only in odd-parity
multiband superconductors with time-reversal symmetry. We show that
the FE pairing potential is odd under the parity operation. To this
end, we start by examining the parity and time-reversal symmetry properties
of the normal state and the pairing potential. A fermionic state $|{\bf k},m_{j}\rangle$,
with ${\bf k}$ as the 3D momentum, and $m_{j}$ being the magnetic
quantum number of total angular momentum $j$, transforms under inversion
$P$ and time-reversal $T$ operations, respectively, as
\begin{eqnarray}
P|{\bf k},m_{j}\rangle & = & |-{\bf k},m_{j}\rangle,\\
T|{\bf k},m_{j}\rangle & = & (-1)^{j+m_{j}}|-{\bf k},-m_{j}\rangle.
\end{eqnarray}
 Inversion symmetry acts only on the momenta and the magnetic quantum
number remains intact. In this case, the matrix form of the inversion
operator is a $4\times4$ identity matrix. The matrix representation
of the anti-unitary time-reversal operator takes the form $\hat{T}=\hat{\mathcal{R}}\mathcal{K}$
with $\mathcal{K}$ being the complex conjugation and the unitary
part $\hat{\mathcal{R}}=e^{i\pi\hat{J}_{y}}=i\hat{\sigma}_{x}\otimes\hat{\sigma}_{y}$.

The normal state preserves inversion and time-reversal symmetries
described by
\begin{align}
\hat{P}\hat{\mathcal{H}}_{{\bf k}}\hat{P}^{-1} & =\hat{\mathcal{H}}_{-{\bf k}},\ \ \ \hat{\mathcal{R}}\hat{\mathcal{H}}_{{\bf k}}^{*}\hat{\mathcal{R}}^{-1}=\hat{\mathcal{H}}_{-{\bf k}}.
\end{align}
 Moreover, the normal state preserves the combination of these symmetries
given by $(\hat{P}\hat{T})\hat{\mathcal{H}}_{{\bf k}}(\hat{P}\hat{T})^{-1}=\hat{\mathcal{H}}_{{\bf k}}$
leading to doubly degenerate eigenstates.

The parity of a pairing potential is distinguished by the orbital
angular momentum of Cooper pairing, i.e., $L=2n+1$ $(L=2n)$ denotes
odd- (even-) parity angular momenta with $n$ being non-negative integers.
In this case, the pairing potential is odd $u$ (even $g$) under
the parity operation given by 
\begin{equation}
\hat{P}\hat{\Delta}_{{\bf k}}\hat{P}^{-1}=-(+)\hat{\Delta}_{-{\bf k}}.
\end{equation}
We can show that the BdG Hamiltonian, given in Eq. (1) of the main
paper, satisfies inversion symmetry for odd- (even-) parity pairing
potential given by 
\begin{equation}
\hat{\mathcal{P}}_{u,g}\hat{H}({\bf k})\hat{\mathcal{P}}_{u,g}^{-1}=\hat{H}(-{\bf k}),
\end{equation}
where $\hat{\mathcal{P}}_{u}=\hat{\sigma}_{z}\otimes\hat{P}$ and
$\hat{\mathcal{P}}_{g}=\hat{\sigma}_{0}\otimes\hat{P}$. Furthermore,
a pairing potential is time-reversal symmetric described by
\begin{align}
\hat{\mathcal{R}}\hat{\Delta}_{{\bf k}}^{*}\hat{\mathcal{R}}^{-1} & =\hat{\Delta}_{-{\bf k}}=\pm\hat{\Delta}_{{\bf k}},
\end{align}
where $-(+)$ corresponds to odd- (even-) parity pairing potentials.
In our investigations, we focus on odd-parity pairing potentials which
preserve time-reversal symmetry.

Due to inversion symmetry in the normal state, the pseudospin operator
$\hat{V}_{{\bf k}}=\{\hat{V}_{{\bf k}}^{+},\hat{V}_{{\bf k}}^{-}\}$
is an even-parity matrix function, i.e., $\hat{V}_{{\bf k}}^{\pm}=\hat{V}_{-{\bf k}}^{\pm}$.
It is given explicitly by

\begin{align}
\hat{V}_{{\bf k}}^{+} & =\frac{1}{2}\frac{|{\bf k}_{||}|}{|\mathbf{k}|}\left(\begin{array}{cc}
2k_{z}k_{-}/k_{+}^{2} & k_{-}/k_{+}\\
\sqrt{3}k_{-}/k_{+} & 0\\
0 & \sqrt{3}\\
1 & -2k_{z}/k_{-}
\end{array}\right),
\end{align}
and 
\begin{align}
\hat{V}_{{\bf k}}^{-} & =\frac{1}{\Gamma_{{\bf k}}^{-}}\left(\begin{array}{cc}
2\sqrt{3}k_{z}k_{-}{\bf k}_{||}^{2}/k_{+}^{2} & -\sqrt{3}k_{-}^{2}\\
-(|\mathbf{k}|^{2}+3k_{z}^{2}){\bf k}_{||}^{2}/k_{+}^{2} & 0\\
0 & |\mathbf{k}|^{2}+3k_{z}^{2}\\
\sqrt{3}{\bf k}_{||}^{2} & 2\sqrt{3}k_{z}k_{+}
\end{array}\right),
\end{align}
where ${\bf k}_{||}^{2}=k_{x}^{2}+k_{y}^{2}$, $\Gamma_{{\bf k}}^{-}=2|\mathbf{k}|\sqrt{|\mathbf{k}|^{2}+3k_{z}^{2}}$,
$|\mathbf{k}|=\sqrt{{\bf k}_{||}^{2}+k_{z}^{2}}$, and $k_{\pm}=k_{x}\pm ik_{y}$.

The parity of the pairing potential projected onto the pseudospin
basis, i.e., $\hat{\Delta}_{{\bf k}}^{\nu\nu^{\prime}}$ with $\nu\nu^{\prime}\in\{+,-\}$,
is the same as the parity of the unprojected pairing potential. We
prove this by the symmetry relation
\begin{align}
\hat{\mathfrak{P}}\hat{\Delta}_{{\bf k}}^{\nu\nu^{\prime}}\hat{\mathfrak{P}}^{-1} & =\hat{\mathfrak{P}}\hat{V}_{{\bf k}}^{\nu\dagger}\hat{\Delta}_{{\bf k}}(\hat{V}_{-{\bf k}}^{\nu^{\prime}\dagger})^{T}\hat{\mathfrak{P}}^{-1}\nonumber \\
 & =\hat{\mathfrak{P}}\hat{V}_{{\bf k}}^{\nu\dagger}(\pm\hat{P}^{-1}\hat{\Delta}_{-{\bf k}}\hat{P})(\hat{V}_{-{\bf k}}^{\nu^{\prime}\dagger})^{T}\hat{\mathfrak{P}}^{-1}\nonumber \\
 & =\pm\hat{V}_{-{\bf k}}^{\nu\dagger}\hat{\Delta}_{-{\bf k}}(\hat{V}_{+{\bf k}}^{\nu^{\prime}\dagger})^{T}\nonumber \\
 & =\pm\hat{V}_{{\bf k}}^{\nu\dagger}\hat{\Delta}_{-{\bf k}}(\hat{V}_{-{\bf k}}^{\nu^{\prime}\dagger})^{T}\nonumber \\
 & =\pm\hat{\Delta}_{-{\bf k}}^{\nu\nu^{\prime}},\label{InversionEigenvector}
\end{align}
where $\hat{\mathfrak{P}}$ is the inversion operator defined by a
$2\times2$ identity matrix in pseudospin basis. Note that to derive
Eq. $\text{\text{\eqref{InversionEigenvector}}}$, we consider the
even-parity property of the eigenspinors. This shows that the inversion
symmetry of the normal state is crucial to the determine the parity
of the projected pairing. Therefore, the FE pairing potential, which
is responsible for TPTs at FEs, has odd- (even-) parity if the orbital
angular momentum of Cooper pairing is odd (even), i.e., 
\begin{equation}
\hat{\mathfrak{P}}\hat{\Delta}_{{\bf k}}^{+-}\hat{\mathfrak{P}}^{-1}=-(+)\hat{\Delta}_{-{\bf k}}^{+-}.
\end{equation}
In the lattice representation, we replace even (odd) momenta in the
finite energy pairing potential $\hat{\Delta}_{k_{\nu}}^{+-}$ by
\begin{align}
\!\! & \!k_{\nu}^{2n+1}\!\rightarrow\![2(1-\text{cos}(k_{\nu}))]^{n}\text{sin}(k_{\nu}),\! & \!\!\text{for }L & =\text{odd},\!\!\label{Odd momenta}\\
\!\! & \!k_{\nu}^{2n}\!\rightarrow\![2(1-\text{cos}(k_{\nu}))]^{n},\! & \!\!\text{for }L & =\text{even},\!\!\label{Even momenta}
\end{align}
where $\nu\in\{x,y,z\}$ and we assume the lattice constant to be
unity. The odd-parity pairing channels contain odd powers of momenta
in $\hat{\Delta}_{{\bf k}}^{+-}$ along the TPTs direction. Thus,
the $\text{sin}(k_{\nu})$ term in Eq. $\text{\text{\eqref{Odd momenta}}}$
forces $\hat{\Delta}_{k_{\nu}=\pm\pi}^{+-}=0$ at the parity-time-reversal-invariant
momenta (PTRIM), i. e., $k_{\nu}=\pm\pi$. This allows to close and
reopen the gap, through the manipulation of the normal state parameters
at FEs. Therefore, helical Dirac surface states emerge only for the
odd-parity pairing channels.

In contrast, bulk bands for even-parity pairing channels at FEs never
close at $k_{\nu}=\pm\pi$ , i.e., $\hat{\Delta}_{k_{\nu}=\pm\pi}^{+-}=4^{n}\neq0$.
This prohibits the quantization of the topological index. Therefore,
the system remains in the topologically trivial phase.

We illustrate the aforementioned general argument by two examples.
Since we focus on the $p$-wave pairing channel, the momenta appear
in linear order, i.e., $L=1$ ($n=0$). Consider the $A_{2u}$ pairing
along the direction ${\bf k}=(0,0,k_{z})$. The BdG Hamiltonian is
block diagonal through the unitary transformation constructed from
$\mathcal{\hat{D}}_{\mathbf{n}}(\pi,\epsilon)$ symmetry in the interband
basis, i.e., $\hat{\mathscr{H}}^{\prime\prime}(0,0,k_{z})=\text{diag}(\hat{h}_{k_{z},+}^{-+},\hat{h}_{k_{z},+}^{+-},\hat{h}_{k_{z},-}^{+-},\hat{h}_{k_{z},-}^{-+})$.
The explicit matrix form of $\hat{h}_{k_{z},\pm}^{+-}$ in the lattice
representation is
\begin{equation}
\hat{h}_{k_{z},\pm}^{+-}=\left(\begin{array}{cc}
-E_{k_{z}}^{-} & \Delta\text{sin}(k_{z})\\
\Delta\text{sin}(k_{z}) & E_{k_{z}}^{+}
\end{array}\right),\label{TPT spectrum}
\end{equation}
where the off-diagonal terms correspond to the FE pairing. In Eq.
$\text{\eqref{TPT spectrum}}$, $-E_{k_{z}}^{-}$ ($E_{k_{z}}^{+}$)
is the hole (electron) band given by
\begin{align}
E_{k_{z}}^{-} & =2(\alpha+\beta/4)(1-\text{cos}(k_{z}))-\mu,\\
E_{k_{z}}^{+} & =2(\alpha+9\beta/4)(1-\text{cos}(k_{z}))-\mu.
\end{align}
The TPT occurs at the PTRIM ${\bf k}=(0,0,\pm\pi)$ where the spectrum
of Eq. $\text{\text{\eqref{TPT spectrum}}}$ becomes
\begin{align}
\varepsilon_{1} & =\mu-(4\alpha+\beta),\ \ \varepsilon_{2}=4\alpha+9\beta-\mu.
\end{align}
We assume the normal state energy bands to have identical sign of
curvature, i.e., $\text{sgn}(\alpha)=\text{sgn}(\beta)=\pm1$. In
this case, the system exhibits a gap at ${\bf k}=(0,0,\pm\pi)$, i.e.,
$|\varepsilon_{1}-\varepsilon_{2}|=2|4a+5b-\mu|>0$. The TPT occurs
when $|\varepsilon_{1}-\varepsilon_{2}|=0$. This results in the phase
transition relation $\mu=4\alpha+5\beta$. Hence, the TPT at nonzero
excitation energies can occur through the interplay between odd-parity
superconductivity and normal state crossings at finite excitation
energies.

In realistic materials, $\alpha$ and $\beta$ are fixed and the system
can exhibit a topological phase for a certain range of chemical potentials.
This is illustrated by the topological phase diagram in Fig. 1(c)
in the main text.

In contrast, the FE pairing never vanishes at PTRIM for even-parity
pairing channels. To illustrate this, consider the even-parity s-wave
$E_{g}$ pairing given by the pairing matrix $\hat{\Delta}_{{\bf k}}=(\Delta/3)(2\hat{J}_{z}^{2}-\hat{J}_{x}^{2}-\hat{J}_{y}^{2})\hat{\mathcal{R}}$
\citep{2DNodes}. This pairing matrix exhibits vanishing (nonvanishing)
intra- (inter-) band pairing at momenta $2k_{z}^{2}=k_{x}^{2}+k_{y}^{2}$,
i.e., $\hat{\Delta}_{{\bf k}}^{\text{intra}}=0$ and $\hat{\Delta}_{{\bf k}}^{+-}\neq0$.
Therefore, this instability channel is a candidate for the TPTs at
FEs. The FE pairing is even under the parity exchange, i.e., $\hat{\mathfrak{P}}\hat{\Delta}_{{\bf k}}^{+-}\hat{\mathfrak{P}}^{-1}=\hat{\Delta}_{-{\bf k}}^{+-}$
with $\hat{\Delta}_{{\bf k}}^{+-}=\Delta(k_{+}/k_{-})\hat{\tau}_{x}$
where $k_{\pm}=k_{x}\pm ik_{y}$. In this case, the interband superconducting
Hamiltonian becomes
\begin{figure}
\centering{}\includegraphics[scale=0.365]{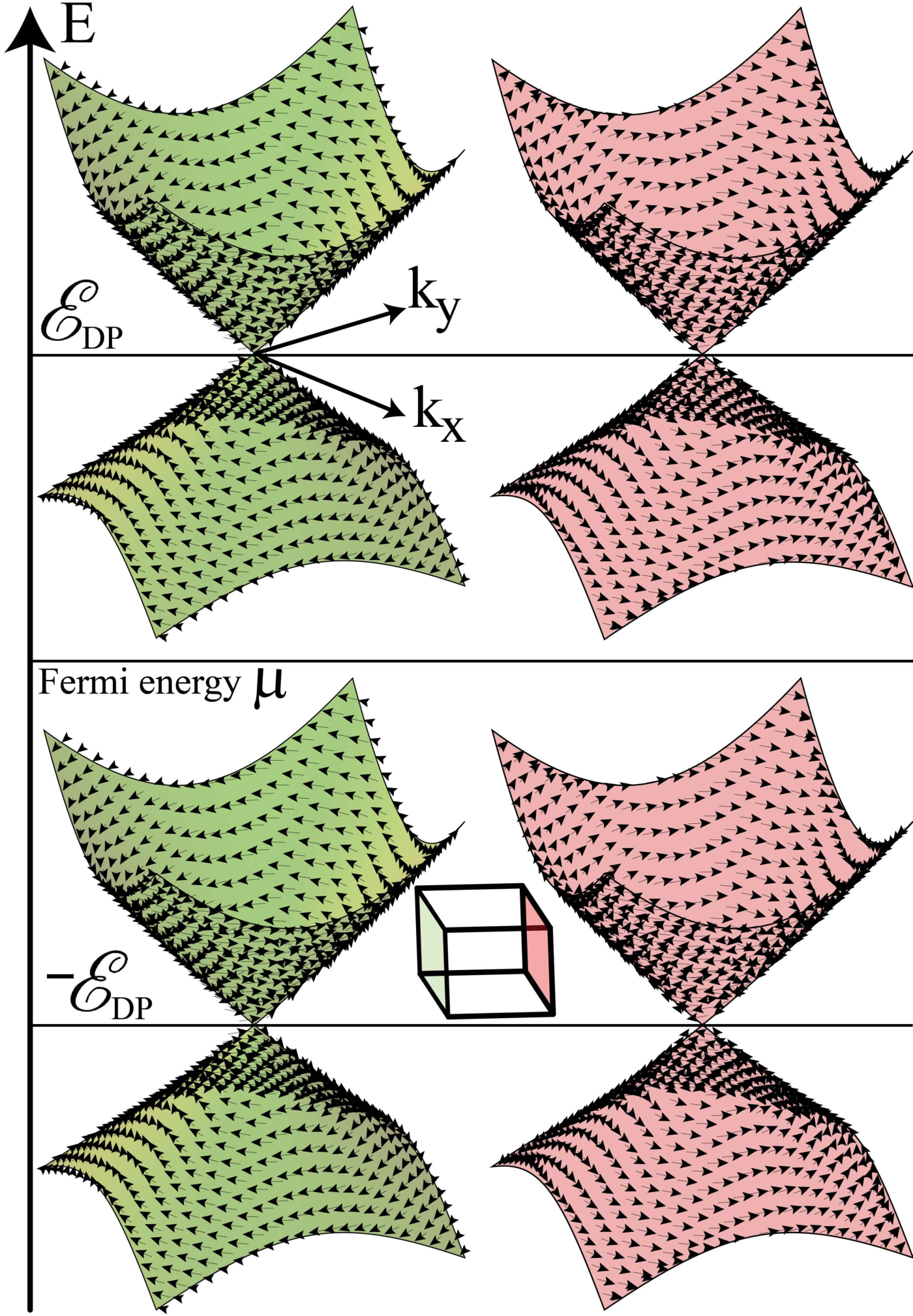}\caption{\label{Helical Spin texture}Helical topological Dirac cones at finite
excitation energies induced by FE Cooper pairing. A pair of helical
Dirac cones (pink) emerge on a surface and the degenerate partner
appears on the other surface (green). Black arrows illustrate the
spin texture of the surface states exhibiting spin-momentum locking
due to the helical property. $\mathscr{E}_{\text{DP}}$ denotes energy
of the Dirac points and $\mu$ is the Fermi energy of the normal state.}
\end{figure}
\begin{equation}
\hat{\mathsf{H}}_{{\bf k}}^{+-}=\left(\begin{array}{cc}
E_{{\bf k}}^{+}\hat{\tau}_{0} & \Delta(k_{+}/k_{-})\hat{\tau}_{x}\\
\Delta(k_{-}/k_{+})\hat{\tau}_{x} & -E_{{\bf k}}^{-}\hat{\tau}_{0}
\end{array}\right),\label{EvenParity}
\end{equation}
where 
\begin{align}
E_{{\bf k}}^{+} & =(3/8)(4\alpha+9\beta)(k_{x}^{2}+k_{y}^{2})-\mu,\\
E_{{\bf k}}^{-} & =(3/8)(4\alpha+\beta)(k_{x}^{2}+k_{y}^{2})-\mu.
\end{align}
Equation $\text{\eqref{EvenParity}}$ preserves a rotational symmetry
along the z-axis in pseudospin basis with rotation angle $\theta=\pi$
given by $\mathcal{\hat{D}}_{\mathbf{n}_{z}}(\pi,1)=\text{diag}(i\hat{\tau}_{z},-i\hat{\tau}_{z})$.
Therefore, $\hat{\mathsf{H}}_{{\bf k}}^{+-}$ becomes diagonal in
the eigenbasis of $\mathcal{\hat{D}}_{\mathbf{n}_{z}}(\pi,1)$ labeled
by eigenvalues $\lambda=\pm i$ obtained by $\hat{\mathcal{U}}^{\dagger}\hat{\mathsf{H}}_{{\bf k}}^{+-}\hat{\mathcal{U}}=\text{diag}(\hat{h}_{{\bf k},+i}^{+-},\hat{h}_{{\bf k},-i}^{+-})$
with
\begin{equation}
\hat{h}_{{\bf k},\pm i}^{+-}=\left(\begin{array}{cc}
-E_{{\bf k}}^{-} & \Delta(k_{-}/k_{+})\\
\Delta(k_{+}/k_{-}) & E_{{\bf k}}^{+}
\end{array}\right),\label{Even-parity Pseuodspin}
\end{equation}
where the matrix of eigenvectors for $\mathcal{\hat{D}}_{\mathbf{n}_{z}}(\pi,1)$
is given by
\begin{equation}
\hat{\mathcal{U}}=\left(\begin{array}{cccc}
0 & 1 & 0 & 0\\
0 & 0 & 0 & 1\\
0 & 0 & 1 & 0\\
1 & 0 & 0 & 0
\end{array}\right).
\end{equation}
The spectrum of Eq. $\text{\eqref{Even-parity Pseuodspin}}$ is always
gapped at FEs. Hence, TPTs at FEs are not possible.

\section{\label{sec:Pseudospin-rotation-symmetry}Pseudospin rotation symmetry
properties}

In this section, we explain the constraint imposed by pseudospin rotation
symmetry on the FE pairing potential. The motivation behind such analysis
is to show the explicit form of a general FE pairing matrix fulfilling
the pseudospin rotation symmetry. To this end, we present the FE pairing
matrix in a general form
\begin{equation}
\hat{\Delta}_{{\bf k}}^{+-}\equiv\boldsymbol{\mathfrak{g}}_{{\bf k}}\cdot\hat{\boldsymbol{\tau}},\label{Spanned Interband}
\end{equation}
where the four-component $\boldsymbol{\mathfrak{g}}_{{\bf k}}$ vector
is given by $\boldsymbol{\mathfrak{g}}_{{\bf k}}=(\mathfrak{g}_{0},\mathfrak{g}_{x},\mathfrak{g}_{y},\mathfrak{g}_{z})$
with the components being momentum dependent functions (this dependency
is dropped). The four-component vector of Pauli matrices is defined
in the interband basis by $\hat{\boldsymbol{\tau}}=(\hat{\tau}_{0},\hat{\tau}_{x},\hat{\tau}_{y},\hat{\tau}_{z})$
with $\hat{\tau}_{0}$ being a $2\times2$ identity matrix.

The matrix representation of FE pairing potential converts to a pairing
function $\delta_{{\bf k},\lambda}^{+-}$ (see Eq. (5) in the main
text) in the presence of $\mathcal{\hat{D}}_{\mathbf{n}}(\pi,\epsilon)$
symmetry where the rotation angle is about the arbitrary unit vector
$\mathbf{n}=(n_{x},n_{y},n_{z})$. In this case, the relation between
$\hat{\Delta}_{{\bf k}}^{+-}$and $\delta_{{\bf k},\lambda}^{+-}$
can be distinguished by the components of the vector $\boldsymbol{\mathfrak{g}}_{{\bf k}}$.
Interestingly, the pseudospin rotation symmetry about the x(y){[}z{]}
axis, defined by 
\begin{equation}
\mathbf{n}_{x}\equiv(1,0,0),\ \mathbf{n}_{y}\equiv(0,1,0),\ \mathbf{n}_{z}\equiv(0,0,1),
\end{equation}
 enforces $\delta_{{\bf k},\lambda}^{+-}$ to select up to two components
of the vector $\boldsymbol{\mathfrak{g}}_{{\bf k}}$. After some algebra,
such constraints for the $\epsilon=-1$ representation can be obtained
by 
\begin{align}
\!\![\mathcal{\hat{D}}_{\mathbf{n}_{z}}(\pi,-1),\hat{\mathsf{H}}_{{\bf k}}^{+-}] & \!=\!0\rightarrow\delta_{{\bf k},\lambda}^{+-}=\mathfrak{g}_{0}+\lambda\mathfrak{g}_{z},\label{nz}\\
\!\![\mathcal{\hat{D}}_{\mathbf{n}_{y}}(\pi,-1),\hat{\mathsf{H}}_{{\bf k}}^{+-}] & \!=\!0\rightarrow\delta_{{\bf k},\lambda}^{+-}=\lambda i\mathfrak{g}_{x}-\mathfrak{g}_{z},\label{ny}\\
\!\![\mathcal{\hat{D}}_{\mathbf{n}_{x}}(\pi,-1),\hat{\mathsf{H}}_{{\bf k}}^{+-}] & \!=\!0\rightarrow\delta_{{\bf k},\lambda}^{+-}=\mathfrak{g}_{0}+\lambda\mathfrak{g}_{x}.\label{nx}
\end{align}
To interpret the above relations, we consider Eq. $\text{\eqref{nz}}$
as an example. In this case, $\delta_{{\bf k},\lambda}^{+-}$ preserves
the pseudospin rotation symmetry under the rotation angle $\theta=\pi$
about the $z$-axis if either $\mathfrak{g}_{0}$ or $\mathfrak{g}_{z}$,
or both of them are finite, and the other components vanish, i.e.,
$\boldsymbol{\mathfrak{g}}_{{\bf k}}=(\mathfrak{g}_{0},0,0,\mathfrak{g}_{z})$.
This is a useful result since we can directly ascertain the pseudospin
rotation symmetry of a FE pairing potential in multiband superconductors
by checking only the components of the vector $\boldsymbol{\mathfrak{g}}_{{\bf k}}$
.

The dot product of the vector $\boldsymbol{\mathfrak{g}}_{{\bf k}}$
and the Pauli matrices in Eq. $\text{\eqref{Spanned Interband}}$
allows us to find the relation between $\hat{\Delta}_{{\bf k}}^{+-}$
and a proper $\mathcal{\hat{D}}_{\mathbf{n}}(\pi,\epsilon)$ symmetry
given by
\begin{align}
\!\!\mathcal{\hat{D}}_{\mathbf{n}_{z}}(\pi,-1): & \ \hat{\Delta}_{{\bf k}}^{+-}=\mathfrak{g}_{0}\hat{\tau}_{0}+\mathfrak{g}_{z}\hat{\tau}_{z},\label{FE pairing 1}\\
\!\!\mathcal{\hat{D}}_{\mathbf{n}_{y}}(\pi,-1): & \ \hat{\Delta}_{{\bf k}}^{+-}=\mathfrak{g}_{x}\hat{\tau}_{x}+\mathfrak{g}_{z}\hat{\tau}_{z},\label{FE pairing 2}\\
\!\!\mathcal{\hat{D}}_{\mathbf{n}_{x}}(\pi,-1): & \ \hat{\Delta}_{{\bf k}}^{+-}=\mathfrak{g}_{0}\hat{\tau}_{0}+\mathfrak{g}_{x}\hat{\tau}_{x}.\label{FE pairing 3}
\end{align}

Importantly, we observe that the component $\mathfrak{g}_{y}$ in
Eqs. (\ref{nz}-\ref{FE pairing 3}) is absent. However, the pseudospin
rotation symmetry with $\epsilon=+1$ representation reveals the allowed
symmetry form of $\delta_{{\bf k},\lambda}^{+-}$ including the $y$-component
of $\boldsymbol{\mathfrak{g}}_{{\bf k}}$,
\begin{align}
\!\![\mathcal{\hat{D}}_{\mathbf{n}_{z}}(\pi,1),\hat{\mathsf{H}}_{{\bf k}}^{+-}] & \!=\!0\rightarrow\delta_{{\bf k},\lambda}^{+-}=\mathfrak{g}_{x}-i\lambda\mathfrak{g}_{y},\\
\!\![\mathcal{\hat{D}}_{\mathbf{n}_{y}}(\pi,1),\hat{\mathsf{H}}_{{\bf k}}^{+-}] & \!=\!0\rightarrow\delta_{{\bf k},\lambda}^{+-}=\mathfrak{g}_{0}+\lambda\mathfrak{g}_{y},\\
\!\![\mathcal{\hat{D}}_{\mathbf{n}_{x}}(\pi,1),\hat{\mathsf{H}}_{{\bf k}}^{+-}] & \!=\!0\rightarrow\delta_{{\bf k},\lambda}^{+-}=-i\lambda\mathfrak{g}_{y}-\mathfrak{g}_{z}.
\end{align}
In this case, the $\mathcal{\hat{D}}_{\mathbf{n}}(\pi,\epsilon)$
symmetry forces the FE pairing matrix to have the explicit form
\begin{align}
\!\!\mathcal{\hat{D}}_{\mathbf{n}_{z}}(\pi,1) & :\ \hat{\Delta}_{{\bf k}}^{+-}=\mathfrak{g}_{x}\hat{\tau}_{x}+\mathfrak{g}_{y}\hat{\tau}_{y},\label{FE pairing 4}\\
\!\!\mathcal{\hat{D}}_{\mathbf{n}_{y}}(\pi,1) & :\ \hat{\Delta}_{{\bf k}}^{+-}=\mathfrak{g}_{0}\hat{\tau}_{0}+\mathfrak{g}_{y}\hat{\tau}_{y},\label{FE pairing 5}\\
\!\!\mathcal{\hat{D}}_{\mathbf{n}_{x}}(\pi,1) & :\ \hat{\Delta}_{{\bf k}}^{+-}=\mathfrak{g}_{y}\hat{\tau}_{y}+\mathfrak{g}_{z}\hat{\tau}_{z}.\label{FE pairing 6}
\end{align}
Taking into account Eqs. (\ref{FE pairing 1}-\ref{FE pairing 3})
and (\ref{FE pairing 4}-\ref{FE pairing 6}), we can understand the
explicit form of the $\mathcal{\hat{D}}_{\mathbf{n}}(\pi,\epsilon)$
symmetry along the TPT direction by looking at the components of $\hat{\Delta}_{{\bf k}}^{+-}$,
e.g., the last column of Table. I in the main text. For instance,
one of the TPT directions for the $T_{2u}$ irrep is $\mathbf{k}\in(\pm k,0,\pm k)$
and the FE pairing potential is given by $\hat{\Delta}_{k}^{+-}=\Delta k(\pm\hat{\tau}_{0}/3-i\hat{\tau}_{y})$.
According to Eq. $\text{\eqref{FE pairing 5}}$, $\hat{\Delta}_{k}^{+-}$
preserves $\mathcal{\hat{D}}_{\mathbf{n}_{y}}(\pi,1)$ symmetry with
$\mathfrak{g}_{0}=\pm\Delta k/3$ and $\mathfrak{g}_{y}=\pm i\Delta k$.

Note that when $\hat{\Delta}_{{\bf k}}^{+-}$ contains only one component
of $\boldsymbol{\mathfrak{g}}_{{\bf k}}$ along the TPT direction,
the rotation axis of the $\mathcal{\hat{D}}_{\mathbf{n}}(\pi,\epsilon)$
is not unique. Importantly, the momentum dependency of components
of the $\boldsymbol{\mathfrak{g}}_{{\bf k}}$ along the TPT direction
is proportional to the orbital angular momentum of Cooper pairing.
For instance, $\boldsymbol{\mathfrak{g}}_{{\bf k}}$ is proportional
to linear order of momenta at TPT directions since we focus on $p$-wave
pairing, i, e., $L=1$.
\begin{figure}
\centering{}\includegraphics[scale=2.5]{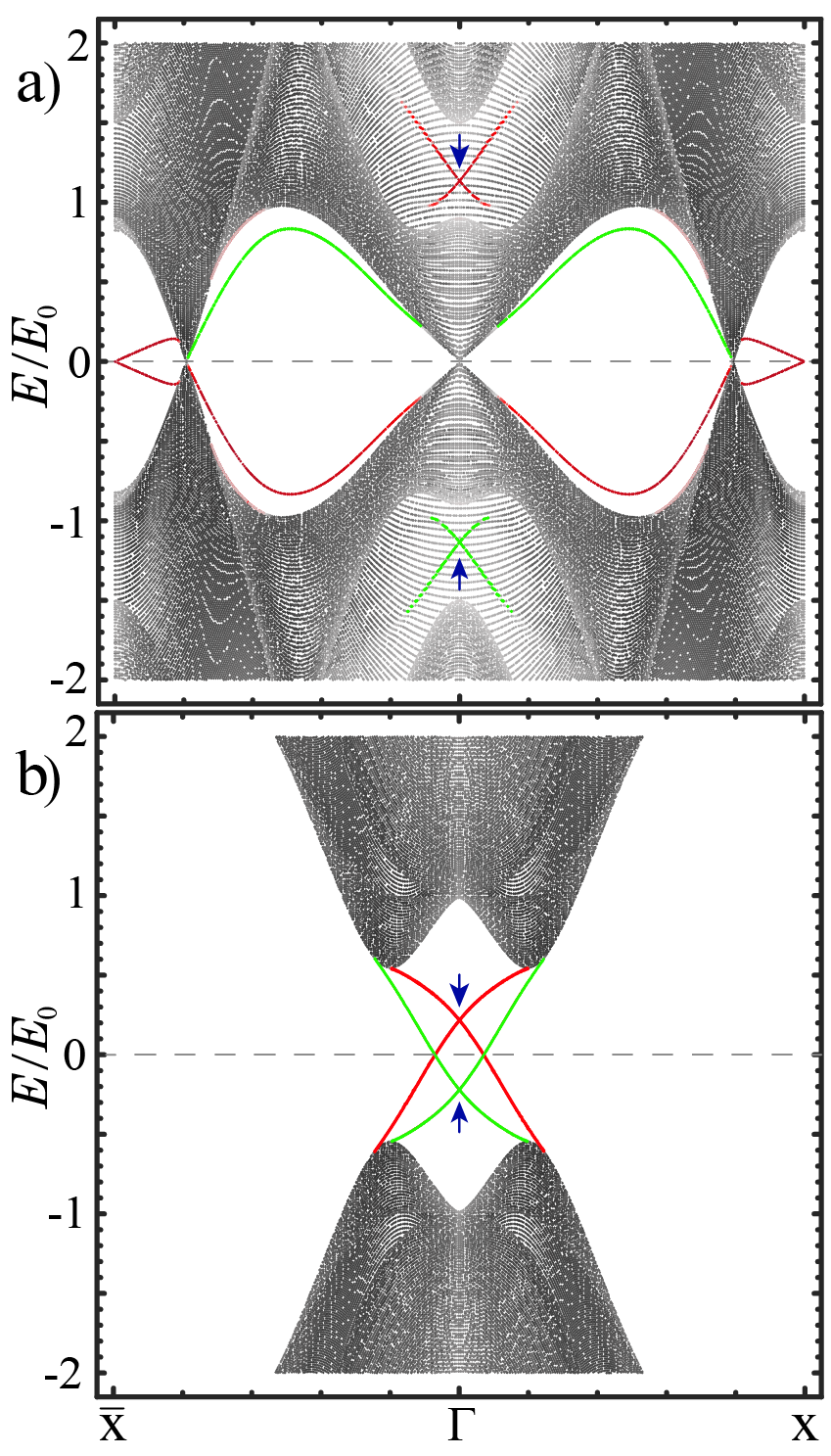}\caption{\label{FigMaj} Enlarged view of the spectra given in Fig. 2(c) of
the main text. The thickness is 180 layers and $E_{0}=\Delta$/2.
The high-symmetry points are $X=(0,\pi/a,0)$ and $\Gamma=(0,0,0)$
where $a$ is the lattice constant in tight binding calculations.
The surface (bulk) states are colorful (dark). Blue arrows indicate
the helical Dirac points induced by topological FE pairing. The gray
dashed line indicates the Fermi energy. (a) {[}(b){]} Surface states
inside (outside) of the bulk states before (after) a Lifshitz transition
for $\mu=-5.2E_{0}$ ($\mu=-E_{0}$). (b) The helical surface states
resemble a butterfly shape, and establish dispersive Majorana modes
away from the high symmetry points at the Fermi energy. The other
parameters are $\alpha=-\Delta/2$ and $\beta=0.3\alpha$.}
\end{figure}

The reason for having two representations for the $\mathcal{\hat{D}}_{\mathbf{n}}(\pi,\epsilon)$
symmetry can be seen by the unitary transformation $\hat{\mathcal{V}}$
between the eigenbases for $\epsilon=+1$ and $\epsilon=-1$,
\begin{equation}
\hat{\phi}_{+1}=\hat{\mathcal{V}}\hat{\phi}_{-1},\ \ \ \hat{\mathcal{V}}=\hat{\mathcal{V}}^{-1}=\hat{\mathcal{V}}^{\dagger}=\left(\begin{array}{cccc}
0 & 0 & 1 & 0\\
0 & 1 & 0 & 0\\
1 & 0 & 0 & 0\\
0 & 0 & 0 & 1
\end{array}\right),
\end{equation}
where $\hat{\phi}_{\epsilon=\pm1}=\hat{\mathcal{U}}^{-1}\hat{\varphi}_{{\bf k}}^{+-}$
with $\hat{\varphi}_{{\bf k}}^{+-}$being the $4\times1$ column as
the basis of FE superconducting Hamiltonian $\hat{\mathsf{H}}_{{\bf k}}^{+-}$,
and $\hat{\mathcal{U}}$ is the eigenvector matrix for $\mathcal{\hat{D}}_{\mathbf{n}}(\pi,\epsilon)$
operator.

\section{Global symmetries\label{sec:Global-symmetries}}

The explicit matrix representations for time-reversal and particle-hole
symmetry operators in BdG formalism are given by $\hat{\mathcal{T}}=i\hat{\varsigma}_{0}\otimes\hat{\sigma}_{x}\otimes\hat{\sigma}_{y}\mathcal{K}$
and $\hat{\mathscr{P}}=\hat{\varsigma}_{x}\otimes\hat{\sigma}_{0}\otimes\hat{\sigma}_{0}\mathcal{K}$,
respectively. Here, $\mathcal{K}$ is the complex-conjugate operator,
$\hat{\varsigma}_{i}$ and $\hat{\sigma}_{i}$ with $i\in\{0,x,y,z\}$
are Pauli matrices acting on particle-hole and spin subspaces, respectively.
The symmetry relation reads $\hat{\mathcal{O}}\hat{H}^{*}({\bf k})\hat{\mathcal{O}}^{-1}=\epsilon\hat{H}(-{\bf k})$
with $\epsilon=+1(-1)$ for $\hat{\mathcal{O}}=\hat{\mathcal{T}}(\hat{\mathscr{P}})$.
Additionally, we can construct a chiral symmetry operator $\hat{\mathcal{C}}=\hat{\mathcal{T}}\hat{\mathscr{P}}$
fulfilling $\hat{\mathcal{C}}\hat{H}({\bf k})\hat{\mathcal{C}}^{-1}=-\hat{H}({\bf k})$
\citep{Topological-Classification-2008,Topological-Classification-2013,Topological-Classification-2016}.
The symmetry operators have the properties $\hat{\mathcal{T}}^{2}=-1$
and $\hat{\mathscr{P}}^{2}=1$. In addition, the inversion symmetry
operator is defined by $\hat{\mathcal{P}}_{u}=\hat{\sigma}_{z}\otimes\hat{P}$
fulfilling $\hat{\mathcal{P}}_{u}^{2}=1$ .

We complete our discussion on topological phase transitions induced
by FE pairing by proposing an alternative approach to calculate the
topological index due to the presence of inversion symmetry \citep{Topological-Classification-2007,Topological-Classification-2011}.
The BdG Hamiltonian commutes with $\hat{\mathcal{P}}_{u}$ and $\hat{T}$
operators at eight PTRIM given by
\begin{align}
\!\!\mathbf{k}_{\text{p}}\! & =\!(0,\!0,\!0),\mathbf{k}_{\text{p}}\!=\!(0,\!0,\!\pi),\mathbf{k}_{\text{p}}\!=\!(0,\!\pi,\!0),\mathbf{k}_{\text{p}}\!=\!(\pi,\!0,\!0),\!\!\nonumber \\
\!\!\mathbf{k}_{\text{p}}\! & =\!(0,\!\pi,\!\pi),\mathbf{k}_{\text{p}}\!=\!(\pi,\!\pi,\!0),\mathbf{k}_{\text{p}}\!=\!(\pi,\!0,\!\pi),\mathbf{k}_{\text{p}}\!=\!(\pi,\!\pi,\!\pi).\!\!\label{PT invariant momenta1}
\end{align}
Generally, the FE pairing arising form odd-parity pairing channels
can induce topological phase transition along the directions $\mathbf{K}$
connecting the PTRIM. The criteria rely on the presence and absence
of finite and low energy pairings along such directions, respectively.
In this case, the BdG Hamiltonian becomes block diagonal
\begin{table}[t]
\begin{centering}
\begin{tabular}{|c|c|c|c|c|}
\hline 
 & $\hat{\mathcal{T}}$ & $\hat{\mathscr{P}}$ & $\hat{\mathcal{C}}$ & $\hat{\mathcal{P}}_{u}$\tabularnewline
\hline 
$\hat{\mathcal{T}}$ & Commute & Commute & Commute & Commute\tabularnewline
\hline 
$\hat{\mathscr{P}}$ &  & Commute & Commute & Anticommute\tabularnewline
\hline 
$\hat{\mathcal{C}}$ &  &  & Commute & Anticommute\tabularnewline
\hline 
$\hat{\mathcal{P}}_{u}$ &  &  &  & Commute\tabularnewline
\hline 
\end{tabular}
\par\end{centering}
\caption{Commutation/anticommutation relations between the discrete symmetries.}
\end{table}
\begin{equation}
\hat{\mathscr{H}}^{\prime}(\mathbf{K})=\text{diag}\big(\hat{h}_{\mathbf{K},+}^{+-},\hat{h}_{\mathbf{K},+}^{+-},\hat{h}_{\mathbf{K},-}^{-+},\hat{h}_{\mathbf{K},-}^{-+}\big),\label{Block}
\end{equation}
where
\begin{equation}
\hat{h}_{\mathbf{K},\lambda}^{+-}=\left(\begin{array}{cc}
-E_{\mathbf{K}}^{-} & \big(\delta_{\mathbf{K},\lambda}^{+-}\big)^{*}\\
\delta_{\mathbf{K},\lambda}^{+-} & E_{\mathbf{K}}^{+}
\end{array}\right).
\end{equation}
In Eq. ($\text{\ref{Block}}$), $\hat{h}_{\mathbf{K},\lambda}^{-+}$
is the particle-hole partner of $\hat{h}_{\mathbf{K},\lambda}^{+-}$
implied by $\hat{\tau}_{y}(\hat{h}_{\mathbf{K},\lambda}^{+-})^{*}\hat{\tau}_{y}^{-1}=\hat{h}_{\mathbf{K},\lambda}^{-+}$.
Note that each block $\hat{h}_{\mathbf{k},\lambda}^{\pm\mp}$ preserves
an effective time-reversal symmetry $\hat{T}=\mathcal{K}$ with $\hat{T}^{2}=+1$.
Particle-hole symmetry and the conventional time-reversal symmetry
with $\hat{T}^{2}=-1$ are broken due to the different diagonal entries
in the symmetry blocks. Also, each block in Eq. ($\text{\ref{Block}}$)
satisfies inversion symmetry implied by $\hat{\tau}_{z}\hat{h}_{\mathbf{K},\lambda}^{\nu\nu^{\prime}}\hat{\tau}_{z}^{-1}=\hat{h}_{-\mathbf{K},\lambda}^{\nu\nu^{\prime}}$
with $\nu,\nu^{\prime}\in\{+,-\}$. In this case, $\hat{h}_{\mathbf{K},\lambda}^{\nu\nu^{\prime}}$
commutes with $\hat{\tau}_{z}$, and the negative parity of the eigenstates
associated to the lower energy band in sector $\hat{h}_{\mathbf{K},\lambda}^{\nu\nu^{\prime}}$
determines the topological nature of the phase transition. Note that
only one of the directions connecting the PTRIM is sufficient to capture
the nontrivial topological phase induced by FE Cooper pairing. Therefore,
to characterize the topological phases, we define a topological index
\begin{equation}
\mathcal{N}=|n_{\Gamma}-n_{\mathbf{p}}|,\label{New toplogical number}
\end{equation}
where $n_{\Gamma}$ and $n_{\mathbf{p}}$ are the number of negative
eigenvalues of the parity operator at the $\Gamma$ point $\mathbf{k}_{\text{p}}=(0,\!0,\!0)$,
and other PTRIM. Note that $\mathcal{N}$ is a $\mathbb{Z}_{2}$ topological
index taking two values, i.e., $\mathcal{N}=1(0)$ in the topologically
nontrivial (trivial) phase. The full topological index can be derived
by the summation of the topological indices associated to the decoupled
blocks in Eq. $\text{\eqref{Block}}$ as
\begin{equation}
Z=\sum_{\lambda=\pm}(\mathcal{N}_{\lambda}^{+-}+\mathcal{N}_{\lambda}^{-+}),\label{Topological index}
\end{equation}
where $\mathcal{N}_{\lambda}^{\nu\nu^{\prime}}$ corresponds to the
block $\hat{h}_{\mathbf{K},\lambda}^{\nu\nu^{\prime}}$. In the topologically
nontrivial phase, conservation of parity leads to the quantization
of the topological index and the emergence of surface states. Pseudospin
rotation symmetry ensures that the surface states come in pairs establishing
FE helical Dirac points. Hence, the appearance of a pair of helical
surface states at positive (negative) excitation energies is signaled
by $\sum_{\lambda=\pm}\mathcal{N}_{\lambda}^{+-}=2$ ($\sum_{\lambda=\pm}\mathcal{N}_{\lambda}^{-+}=2$).
Considering two surfaces, we observe four Dirac surface states on
each surface (two for positive excitation energies and two for negative
energies) due to parity-time-reversal symmetry, cf. Fig. $\text{\ref{Helical Spin texture}}$.

A Lifshitz transition can move the FE helical Dirac surface states,
shown in Fig. $\text{\ref{FigMaj}}$(a), to the low energies. In this
case, the surface states establish Majorana modes at the Fermi energy,
see Fig. $\text{\ref{FigMaj}}$(b).

Note that the point group symmetry of the normal-state Hamiltonian
for any real system is lower than the O(3) symmetry of the Luttinger-Kohn
Hamiltonian discussed in the main text. The Luttinger-Kohn Hamiltonian
serves as a valuable approximation for energy bands close to high-symmetry
points. However, the symmetry of the superconducting (BdG) Hamiltonian
is relevant for our analysis. It belongs to the $O_{h}$ symmetry
group, as the O(3) symmetry of the normal state transforms into cubic
point group symmetry due to the pairing matrices derived through irreducible
representations of the cubic point group symmetry. Importantly, the
pseudospin rotation symmetry mentioned above is present for multiband
system with parity-time-reversal symmetry. Its presence is independent
of the O(3) symmetry of the normal state Hamiltonian. Nonetheless,
we examine our predictions when the normal state is influenced by
O(3) symmetry-breaking terms. To this end, we incorporate the cubic
spin-orbit coupling term to the normal state Hamiltonian as
\begin{figure}
\centering{}\includegraphics[scale=1.84]{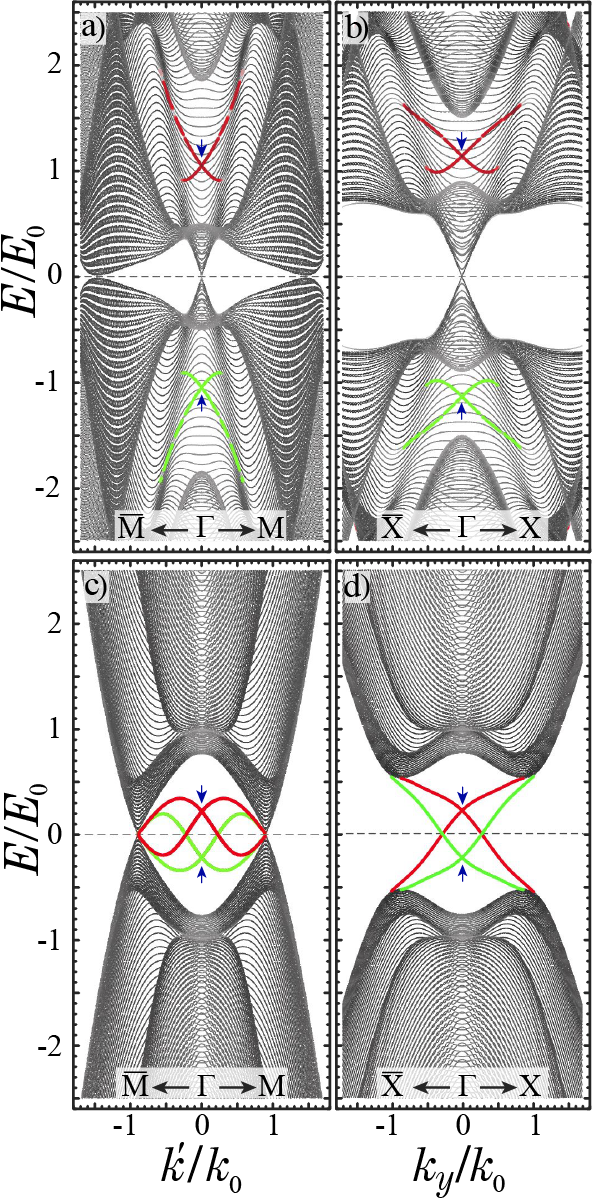}\caption{\label{Fig_cubicTerm}Spectra of a slab with (001) surfaces in the
topologically nontrivial regime. The thickness is 100 layers. The
cubic spin-orbit coupling term is chosen as $\gamma=\alpha$. Hence,
O(3) symmetry is broken in the normal state. In panels (a) and (c),
$k^{\prime}$ defines the direction in momentum space with $k_{x}=k_{y}$.
The chemical potential is chosen as (a) $\mu=-5.2E_{0}$, (b) $\mu=-5.1E_{0}$,
and (c,d) $\mu=-E_{0}$. The other parameters are $\beta=0.3\alpha$,
$\alpha=-(2/3)\Delta$, and $E_{0}=(2/3)\Delta$. The high-symmetry
points are $X=(0,\pi/a,0)$ and $M=(\pi/a,\pi/a,0)$.}
\end{figure}

\begin{equation}
\!\!\mathcal{\hat{H}}({\bf k})\!=\alpha|{\bf k}|^{2}\hat{I}_{4}\!+\!\beta\sum_{i}k_{i}^{2}\hat{J}_{i}^{2}\!+\!\gamma\sum_{i\neq j}k_{i}k_{j}\hat{J}_{i}\hat{J}_{j}-\mu,\!\label{EqCubic}
\end{equation}
where $\gamma$ parametrizes the strength of the cubic spin-orbit
coupling term. This Hamiltonian refers to the O(3) symmetry-broken
case in the normal state if $\gamma\neq\beta$. We adopt the tight-binding
regularization of Eq. ($\ref{EqCubic}$) in the BdG form. The spectral
result within the topologically non-trivial phase is illustrated in
Fig. $\ref{Fig_cubicTerm}$. Evidently, we can also observe the emergence
of helical Dirac surface states with particle-hole character away
form the Fermi energy due to the unconventional finite energy Cooper
pairing in the case with broken O(3) symmetry, see Fig. $\ref{Fig_cubicTerm}$(a,b).
In addition, we investigate a Lifshitz transition around the $\Gamma$
point by a variation of the chemical potential as illustrated in Fig.
$\ref{Fig_cubicTerm}$(c,d). In this case, the surface states are
shifted towards low energies. If the superconducting gap possesses
nodes at zero excitation energy, these surface states connect the
nodal points, see Fig. $\ref{Fig_cubicTerm}$(c). If the superconducting
gap is fully developed at zero excitation energy, these surface states
connect to the bulk states at finite excitation energy, see Fig. $\ref{Fig_cubicTerm}$(d).

\section{Point group symmetry analysis\label{sec:Point-group-symmetry}}

The block diagonalization of the BdG Hamiltonian is possible in directions
where the pairing potential satisfies two conditions:
\begin{enumerate}
\item \label{enu:ITEM1}vanishing intraband pairing,
\item \label{enu:ITEM2}nonvanishing interband pairing at finite energies.
\end{enumerate}
These two conditions can be fulfilled when time-reversal $T$ symmetry
combines with twofold rotation about the $\langle110\rangle$ axis.
This symmetry operator is defined by $\mathcal{\hat{C}}_{T}\equiv\hat{T}\hat{C}_{2,x+y}$
having the property $\mathcal{\hat{C}}_{T}^{2}=1$. We then use the
$\hat{C}_{2,x+y}$ operator combined with inversion $\hat{P}$ symmetry
(mirror reflection), denoted as $\mathcal{\hat{C}}_{P}\equiv\hat{P}\hat{C}_{2,x+y}$,
for block diagonalization. In addition, the double degeneracy of states
at each momenta is ensured by $\hat{P}\hat{T}=e^{-i\pi}\mathcal{\hat{C}}_{T}\mathcal{\hat{C}}_{P}$.
Note that $\hat{T}$, $\hat{C}_{2,x+y}$, and $\hat{P}$ commute with
each other, and $\mathcal{\hat{C}}_{P}^{2}=-1$.

In the following, we illustrate these points by group theoretical
analysis. We begin by analyzing the point group symmetry of the given
BdG Hamiltonian
\begin{equation}
\hat{H}_{\text{BdG}}(\mathbf{k})\!=\!\left(\begin{array}{cc}
\hat{\mathcal{H}}(\mathbf{k}) & \hat{\Delta}(\mathbf{k})\\
\hat{\Delta}^{\dagger}(\mathbf{k}) & -\hat{\mathcal{H}}^{T}(-\mathbf{k})
\end{array}\right)\!\!\!\!\!\label{BdGHmm}
\end{equation}
where $\hat{\mathcal{H}}(\mathbf{k})$ and $\hat{\Delta}(\mathbf{k})=\hat{D}(\mathbf{k})e^{i\pi\hat{J}_{y}}$
denote normal state and pairing matrix, respectively. $\hat{H}_{\text{BdG}}(\mathbf{k})$
preserves the symmetry group $G$ if its symmetry elements, denoted
as $\hat{g}$, satisfy the following condition
\begin{align}
\hat{\mathcal{H}}(\mathbf{k}) & \longmapsto\hat{g}\hat{\mathcal{H}}(R^{-1}\mathbf{k})\hat{g}^{\dagger},\ \ \hat{D}(\mathbf{k})\longmapsto\hat{g}\hat{D}(R^{-1}\mathbf{k})\hat{g}^{\dagger},\label{Pairing under point}
\end{align}
where $\hat{g}$ can for instance be a $q$-fold rotation and $R$
is a $3\times3$ orthogonal matrix implementing rotation on momentum
space. Then, the point group symmetry operation for Eq. $\text{(\ref{BdGHmm})}$
takes the form
\begin{equation}
\hat{\mathcal{G}}\hat{H}_{\text{BdG}}(R^{-1}\mathbf{k})\hat{\mathcal{G}}^{-1}=\hat{H}_{\text{BdG}}\left(\mathbf{k}\right),
\end{equation}
where $\hat{\mathcal{G}}=\text{diag}(\hat{g},\pm\hat{g}^{*})$. 
In our system, $G$ for the superconducting Hamiltonian is given by
\begin{equation}
G=U(1)\otimes\mathscr{P}\otimes T\otimes O_{h},
\end{equation}
where $U(1)$ is a global phase-rotation symmetry, $\mathscr{P}$
($T$) denotes anti-unitary particle-hole (time-reversal) symmetry,
and $O_{h}$ is a cubic point group symmetry. $O_{h}$ describes the
combination of inversion symmetry $P$ and octahedral $O$ point group
symmetry. The $O$ group consists of $q-$fold rotations about the
$\mathbf{n}$ axis labeled by $C_{q,\mathbf{n}}$. Combining $P$
with $O$, this results in $q-$fold improper rotations denoted by
$PC_{q,\mathbf{n}}$.

In our study, the pairing channels are odd (even) under $\hat{P}$
$(\hat{T})$ symmetries such that $\hat{M}\hat{D}(-\mathbf{k})\hat{M}^{\dagger}=-(+)\hat{D}(\mathbf{k})$
with $\hat{M}=\hat{P}(\hat{T})$. Moreover, $\hat{D}(\mathbf{k})$
anti-commute with $\hat{P}\hat{T}$ symmetry, i.e., $\{\hat{D}(\mathbf{k}),\hat{P}\hat{T}\}=0$
with $\hat{T}=\text{exp}(i\pi\hat{J}_{y})\mathcal{K}$ and $\hat{P}=\hat{I}_{4}$,
where $\mathcal{K}$ is complex conjugation. Importantly, the generators
of $\hat{D}(\mathbf{k})$ in combination with $\hat{T}$ symmetry,
this can impose constraints on the pairing potential. In the main
text, we focus on the $A_{2u}$ pairing channel (spin-septet). The
matrix representation of $\hat{D}(\mathbf{k})$ is $\hat{D}(\mathbf{k})=\mathbf{k}\cdot\hat{\mathbf{T}}$
where $\hat{T}_{i}=\{\hat{J}_{i},\hat{J}_{i+1}^{2}-\hat{J}_{i+2}^{2}\}$
with $i+1=y$ if $i=x$, etc., cyclically. The generators for the
$A_{2u}$ channel are $\{e^{i\pi}\hat{C}_{4z},e^{i\pi}\hat{C}_{2,z+x}\}$,
where $\hat{C}_{2,z+x}$ ($\hat{C}_{4z}$) denotes two(four)fold rotation
about the {[}1,0,1{]} ({[}0,0,1{]}) axis \citep{VenderbosPRX}. Note
that $\hat{C}_{4z}$ does not constrain the pairing channel while
the rotation about $C_{2}^{\prime}$ axis does. In this case, normal
state and pairing channel transform under twofold rotation about the
$C_{2}^{\prime}$ axis, e.g., {[}1,1,0{]} axis, such that
\begin{align}
\hat{C}_{2,x+y}\hat{D}(k_{y},k_{x},-k_{z})\hat{C}_{2,x+y}^{\dagger} & =-\hat{D}(\mathbf{k}),\\
\hat{C}_{2,x+y}\hat{\mathcal{E}}(k_{y},k_{x},-k_{z})\hat{C}_{2,x+y}^{\dagger} & =\hat{\mathcal{E}}(\mathbf{k}).
\end{align}
Combining $\hat{T}$ with twofold rotation symmetry denoted by $\mathcal{\hat{C}}_{T}=\hat{T}\hat{C}_{2,x+y}$,
yielding the relation
\begin{align}
\mathcal{\hat{C}}_{T}\hat{D}(-k_{y},-k_{x},k_{z})\mathcal{\hat{C}}_{T}^{\dagger} & =-\hat{D}(\mathbf{k}),\label{CTsymmetry1}\\
\mathcal{\hat{C}}_{T}\hat{\mathcal{E}}(-k_{y},-k_{x},k_{z})\mathcal{\hat{C}}_{T}^{\dagger} & =\hat{\mathcal{E}}(\mathbf{k}).\label{CTsymmetry2}
\end{align}
In this case, $\mathcal{\hat{C}}_{T}$ results in vanishing (nonvanishing)
intraband (interband) pairing potentials.

We prove this at two Fermi momenta $\mathbf{k}_{1}$ and $\mathbf{k}_{2}$
(due to having two energy bands), and the crossing momenta $\mathbf{k}_{3}$
at finite energies. Note that $\mathbf{k}_{1}$ ($\mathbf{k}_{2}$)
is associated to the Fermi momentum for the $|m_{j}|=3/2$ ($|m_{j}|=1/2)$
Fermi surface. The $|m_{j}|=1/2$ Fermi surface is constrained in
a similar way by $\mathcal{\hat{C}}_{T}$ as the $|m_{j}|=3/2$ Fermi
surface. This results in point nodes in both double-degenerate Fermi
surfaces along the $\langle001\rangle$ direction. We prove this through
symmetry analysis in pseudospin (Kramer's partner) representation.
The double-degeneracy is guaranteed by $PT$ symmetry. In this case,
the effective pairing projected onto the intraband basis at $\mathbf{k}_{1}$
can be represented in the pseudospin basis as
\begin{equation}
\hat{\Delta}_{\text{eff}}^{\nu\nu}(\mathbf{k}_{1})\equiv(\mathbf{d}(\mathbf{k}_{1})\cdot\hat{\sigma})(i\hat{\sigma}_{y}),
\end{equation}
where $\mathbf{d}(\mathbf{k}_{1})=(d_{x}(\mathbf{k}_{1}),d_{y}(\mathbf{k}_{1}),d_{z}(\mathbf{k}_{1}))$,
$\hat{\sigma}=(\hat{\sigma}_{x},\hat{\sigma}_{y},\hat{\sigma}_{z})$
is the vector of Pauli matrices in intraband pseudospin basis. Note
that $\hat{\Delta}_{\text{eff}}^{\nu\nu}(\mathbf{k}_{1})$ is represented
in pseudospin triplet state due to the odd parity of $\hat{D}(\mathbf{k})$.
For $\nu=+$, $\hat{\Delta}_{\text{eff}}^{\nu\nu}(\mathbf{k}_{1})$
describes pairing of $|m_{j}|=3/2$ states at the Fermi energy. At
$\mathbf{k}_{1}$, spin and momentum transform under $\mathcal{\hat{C}}_{T}$
as
\begin{equation}
\mathcal{\hat{C}}_{T}\hat{\sigma}\mathcal{\hat{C}}_{T}^{\dagger}=(-\hat{\sigma}_{y},-\hat{\sigma}_{x},\hat{\sigma}_{z}),\ R^{-1}\mathbf{k}_{1}=(-k_{y},-k_{x},k_{z}).\label{CT symmetry}
\end{equation}
Consequently, along the $\langle001\rangle$ direction, the effective
pairing potential becomes $\hat{\Delta}_{\text{eff}}^{++}(\mathbf{k}_{1})=d_{z}(k_{z})\hat{\sigma}_{z}(i\hat{\sigma}_{x})$
which violates the Fermi statistics unless $d_{z}(\mathsf{k}_{z})=0$.
Consequently, this results in point nodes along z-direction (and all
equivalent directions). Moreover, the same holds true for $\hat{\Delta}_{\text{eff}}^{--}(\mathbf{k}_{2})$.
Therefore, point $\text{\ref{enu:ITEM1}}.$, stated at the beginning
of this section, is fulfilled leading to
\begin{equation}
\hat{\Delta}_{\text{eff}}^{++}(\mathbf{k}_{1})=\hat{\Delta}_{\text{eff}}^{--}(\mathbf{k}_{2})=0.\label{VanishingIntraband}
\end{equation}
To realize point $\text{\ref{enu:ITEM2}}$., stated at the beginning
of this section, we analyze how interband pairing is affected by $\mathcal{\hat{C}}_{T}$
symmetry. Finite energy pairing happens at interband momenta $\mathbf{k}_{3}$
where double-degenerate $|m_{j}|=3/2$ electron states cross with
$|m_{j}|=1/2$ hole bands away from the Fermi energy. We can expand
the interband pairing matrix $\hat{\Delta}_{\text{eff}}^{+-}(\mathbf{k}_{3})$
as
\begin{equation}
\hat{\Delta}_{\text{eff}}^{+-}(\mathbf{k}_{3})=\mathfrak{g}_{\mathbf{k}_{3}}^{+-}\cdot\hat{\mathbf{\tau}},
\end{equation}
where $\mathfrak{g}_{\mathbf{k}_{3}}^{+-}$ is a four component vector
given by
\begin{equation}
\mathfrak{g}_{\mathbf{k}_{3}}^{+-}=(\mathrm{\mathfrak{g}}_{0,\mathbf{k}_{3}}^{+-},\mathrm{\mathfrak{g}}_{x,\mathbf{k}_{3}}^{+-},\mathfrak{g}_{y,\mathbf{k}_{3}}^{+-},\mathrm{\mathfrak{g}}_{z,\mathbf{k}_{3}}^{+-}),
\end{equation}
and $\hat{\mathbf{\tau}}=(\hat{\tau}_{0},\hat{\tau}_{x},\hat{\tau}_{y},\hat{\tau}_{z})$
being the four component vector of Pauli matrices represented in interband
pseudospin basis. Applying $\mathcal{\hat{C}}_{T}$ symmetry on $\hat{\mathbf{\tau}}$
produce the same results as mentioned in Eq. ($\text{\ref{CT symmetry}}$)
for $\hat{\sigma}$. In this case, $\hat{\tau}_{0}$ and $\hat{\tau}_{z}$
remain invariant along the $\langle001\rangle$ direction. Consequently,
the interband pairing matrix becomes
\begin{equation}
\hat{\Delta}_{\text{eff}}^{+-}(k_{z})=\mathrm{\mathfrak{g}}_{0,k_{z}}^{+-}\hat{\tau}_{0}+\mathrm{\mathfrak{g}}_{z,k_{z}}^{+-}\hat{\tau}_{z}.\label{NonvanishingIterband}
\end{equation}
It is worth mentioning that $\hat{\tau}_{0}$ is allowed by the Pauli
exclusion principle if we exchange band indices in addition to spin
indices. Importantly, $\hat{\Delta}_{\text{eff}}^{+-}(k_{z})$ is
constrained by $\mathcal{\hat{C}}_{T}$ such that only up to two components
of the $\mathfrak{g}_{\mathbf{k}_{2}}^{+-}$ vector are finite. This
is identical with our analysis based on rotational symmetry in pseudospin
basis described in Sec. $\text{\ref{sec:Pseudospin-rotation-symmetry}}$.
The $\hat{\tau}_{z}$ remains invariant under two-fold rotation in
pseudospin space combined with time-reversal symmetry.

Despite our pairing model is odd in parity, we can define an inversion
operator in BdG form as
\begin{equation}
\hat{\mathcal{G}}=\text{diag}(\hat{I}_{4},-\hat{I}_{4})=\hat{\sigma}_{z}\otimes\hat{I}_{4}
\end{equation}
where
\begin{equation}
\hat{\mathcal{G}}\hat{H}_{\text{BdG}}(-\mathbf{k})\hat{\mathcal{G}}^{-1}=\hat{H}_{\text{BdG}}(\mathbf{k}).
\end{equation}
In this case, the results given in Eqs. $\text{(\ref{VanishingIntraband}})$
and $\text{(\ref{NonvanishingIterband}})$ hold true for $\mathcal{\hat{C}}_{P}$.

Importantly, such symmetries enforce the BdG Hamiltonian to\textit{
}become block diagonal along the topological phase transition directions
and at finite excitation energies. This is a direct consequence of
Eqs. $\text{(\ref{VanishingIntraband}})$ and $\text{(\ref{NonvanishingIterband}})$.
Although $\mathcal{\hat{C}}_{T}$ is anti-unitary, the twofold degeneracy
can be lifted in the eigenspace of mirror reflection symmetry as 
\begin{equation}
\hat{Y}^{-1}\hat{H}_{\text{BdG}}(k_{z})\hat{Y}=\text{diag}\left(\hat{H}_{+i}(k_{z}),\hat{H}_{-i}(k_{z})\right),
\end{equation}
where $\hat{Y}$ is the matrix of eigenvectors for $\mathcal{\hat{C}}_{P}$
operator, and $\hat{H}_{\lambda}(\mathbf{k})$ is a $4\times4$ block
labeled with eigenvalues of $\hat{P}\hat{C}_{2,x+y}$ as $\lambda=\pm i$.
Note that $\hat{H}_{+i}(k_{z})=\hat{H}_{-i}(k_{z})$ with

\begin{equation}
\hat{H}_{+i}(k_{z})=\left(\begin{array}{cccc}
-E_{k_{z}}^{+} & 0 & 0 & \varDelta k_{z}\\
0 & -E_{k_{z}}^{-} & \varDelta k_{z} & 0\\
0 & \varDelta k_{z} & E_{k_{z}}^{+} & 0\\
\varDelta k_{z} & 0 & 0 & E_{k_{z}}^{-}
\end{array}\right),
\end{equation}
where $\varDelta=(\sqrt{3}/2)\Delta$. The pairing sector is situated
on the off-diagonal block of $\hat{H}_{\pm i}(\mathbf{k})$ with vanishing
intraband pairing. $\hat{H}_{\pm i}(\mathbf{k})$ can be further brought
into block diagonal form through the transformation $\hat{W}$ onto
the interband basis.  Such a transformation is given by
\begin{align}
\hat{\mathscr{H}}^{\prime} & (k_{z})=\hat{W}^{-1}\left(\hat{Y}^{-1}\hat{H}_{\text{BdG}}(k_{z})\hat{Y}\right)\hat{W}\nonumber \\
 & =\text{diag}\left(\hat{h}_{k_{z},+i}^{-+},\hat{h}_{k_{z},+i}^{+-},\hat{h}_{k_{z},-i}^{-+},\hat{h}_{k_{z},-i}^{+-}\right),\label{PC2SEigenspace}
\end{align}
where 
\begin{equation}
\hat{h}_{k_{z},\lambda}^{\nu\nu^{\prime}}=\left(\begin{array}{cc}
-E_{k_{z}}^{\nu^{\prime}} & \varDelta k_{z}\\
\varDelta k_{z} & E_{k_{z}}^{\nu}
\end{array}\right),
\end{equation}
and
\begin{equation}
\hat{W}=\left(\begin{array}{cccccccc}
1 & 0 & 0 & 0 & 0 & 0 & 0 & 0\\
0 & 0 & 1 & 0 & 0 & 0 & 0 & 0\\
0 & 0 & 0 & 1 & 0 & 0 & 0 & 0\\
0 & 1 & 0 & 0 & 0 & 0 & 0 & 0\\
0 & 0 & 0 & 0 & 1 & 0 & 0 & 0\\
0 & 0 & 0 & 0 & 0 & 0 & 1 & 0\\
0 & 0 & 0 & 0 & 0 & 0 & 0 & 1\\
0 & 0 & 0 & 0 & 0 & 1 & 0 & 0
\end{array}\right).
\end{equation}
Notably, Eq. $\text{(\ref{PC2SEigenspace}})$ is identical to Eq.
(6) of the main text. The effective Hamiltonian $\hat{\mathtt{H}}(k_{x},k_{y})$
for the helical surface states at finite excitation energies (see
Eq. (8) of the main text) fulfills the symmetry relation
\begin{align}
\mathcal{\hat{C}}_{T}\hat{\mathtt{H}}(k_{x},k_{y})\mathcal{\hat{C}}_{T}^{-1} & =\mathcal{\hat{C}}_{P}\hat{\mathtt{H}}(k_{x},k_{y})\mathcal{\hat{C}}_{P}^{-1}=\hat{\mathtt{H}}(-k_{y},-k_{x}).
\end{align}

\section{Stability of helical Dirac surface states at finite energies\label{sec:Stability-of-helical}}

\begin{figure}
\begin{centering}
\includegraphics[scale=0.915]{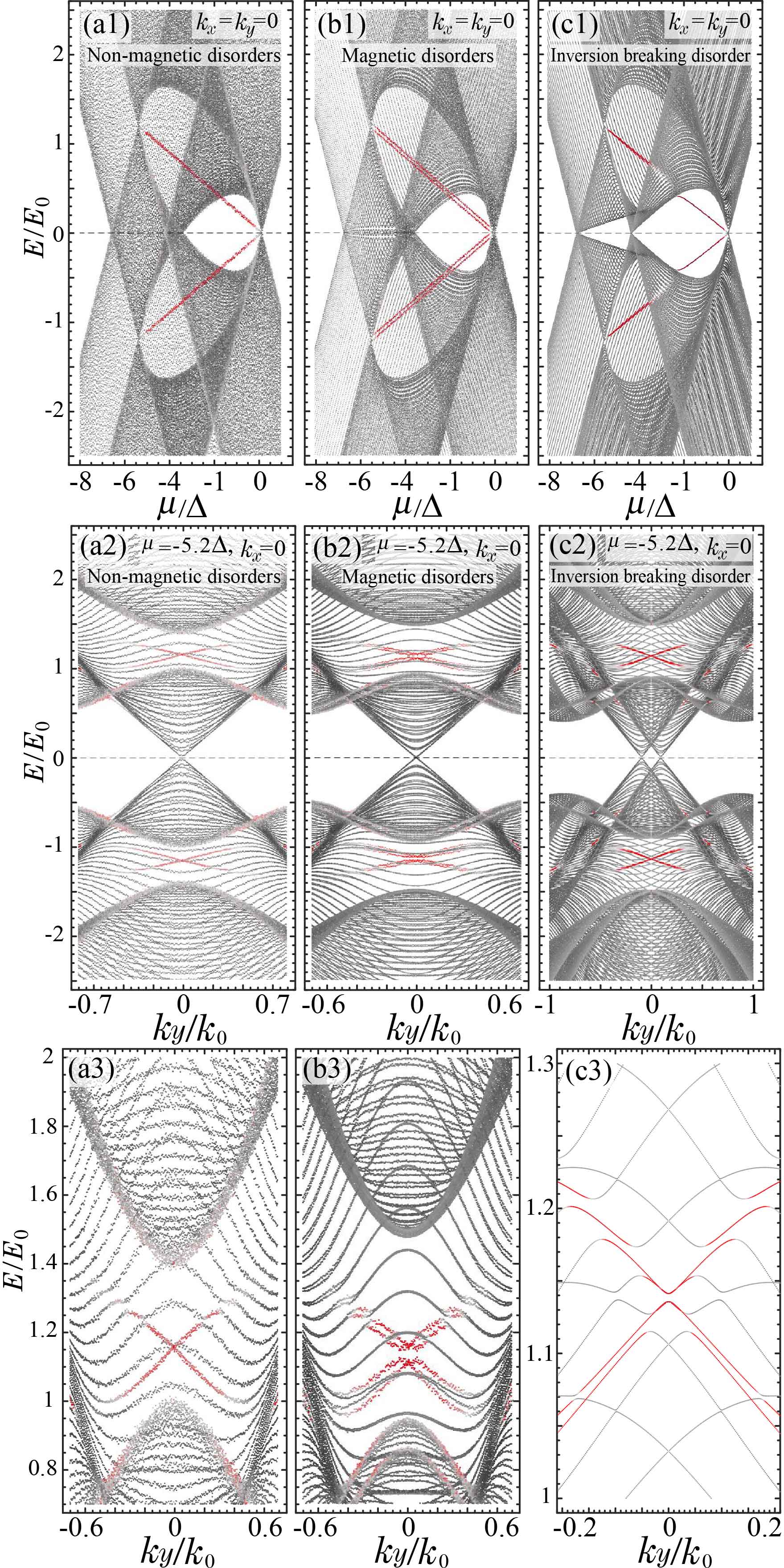}
\par\end{centering}
\caption{\label{Disorders}Excitation spectrum for (001) slab in the presence
of nonmagnetic randomness with strength (a1) $v_{i}\in\{0,\Delta\}$
for 120 layers, (a2,a3) $v_{i}\in\{0,0.6\Delta\}$ for 80 layers,
(b2-b3) magnetic randomness with the form $M_{x,n_{i}}\hat{J}_{x}+M_{y,n_{i}}\hat{J}_{y}$
with $M_{x(y),n_{i}}\in\{0,0.2\Delta\}$ for 80 layers (c1-c3) inversion
symmetry breaking spin-orbit coupling $\delta=0.3\Delta$ for 120
layers. Panels (a3), (b3) and (c3) are the enlarged view of the top
panels. Other parameters are $(\alpha,\beta,\gamma,\mu)=-(1,0.3,1,5.2)\Delta$.
Note that $\gamma\protect\neq\beta$ indicates broken O(3) symmetry
in the normal state. The extended bulk (surface) states are illustrated
by gray (red) color according to inverse participation ratio.}
\end{figure}

To investigate the stability of helical topological surface states
at finite excitation energies, we add randomness to chemical potential
and magnetization of each layer in $z$-direction by
\begin{equation}
H_{R}(\mathbf{k}_{\shortparallel},z)\!=\!\sum_{n_{z}}\sum_{\mathbf{k}_{\shortparallel}}\!\hat{\psi}_{\mathbf{k}_{\shortparallel},z}^{\dagger}\!\left(\mu_{n_{z}}+\mathbf{M}_{n_{z}}\cdot\hat{\mathbf{J}}\right)\!\hat{\psi}_{\mathbf{k}_{\shortparallel},z},\!\!\!\label{Impurities}
\end{equation}
where the basis in $j=3/2$ representation is
\begin{equation}
\hat{\psi}_{\mathbf{k}_{\shortparallel},z}^{\dagger}=(c_{\frac{3}{2},\mathbf{k}_{\shortparallel},z}^{\dagger},c_{\frac{1}{2},\mathbf{k}_{\shortparallel},z}^{\dagger},c_{-\frac{1}{2},\mathbf{k}_{\shortparallel},z}^{\dagger},c_{-\frac{3}{2},\mathbf{k}_{\shortparallel},z}^{\dagger}).
\end{equation}
In Eq. $\text{(\ref{Impurities}})$, we consider $k_{z}$ to be no
longer conserved. Instead, $n_{z}=0,...,N_{z}$ is the layer index
with $N_{z}$ the total number of layers along the z-axis; $\sum_{\mathbf{k}_{\shortparallel}}=\sum_{k_{x}}\sum_{k_{y}}$,
$\mathbf{k}_{\shortparallel}=(k_{x},k_{y})$ represents the conserved
momenta; $\hat{\mathbf{J}}=(\hat{J}_{x},\hat{J}_{y},\hat{J}_{z})$
is the vector of angular momenta; $\mu_{n_{i}}$ denotes the nonmagnetic
onsite potential at layer $n_{i}$ where the strength is a uniformly
distributed random number within the interval $\mu_{n_{i}}\in[0,\Delta]$;
the Zeeman field vector is $\mathbf{M}_{n_{i}}=(M_{x,n_{i}},M_{y,n_{i}},0)$,
with the strength $M_{x(y),n_{i}}$, taken as uniformly distributed
random number in the interval $M_{x(y),n_{i}}\in[0,0.06\Delta]$.
The numerical results are shown in Fig. $\text{\ref{Disorders}}$(a1-a3)
and $\text{\ref{Disorders}}$(b1-b3). We can observe in Fig. $\text{\ref{Disorders}}$(a1)
and (a2) that bulk and surface states become broadened while the topological
surface states remain intact in the presence of nonmagnetic randomness
at finite excitation energies since $\hat{C}_{T}$ symmetry is preserved.
When the system is subjected to magnetic randomness, the degeneracy
of surface states are lifted due to broken time-reversal symmetry.
In this case, $\hat{C}_{T}$ is broken. Along $k_{y}$ direction (see
Fig. $\text{\ref{Disorders}}$(b2) and $\text{\ref{Disorders}}$(b3)),
the Dirac cone is gapped.

Furthermore, the symmetry $\hat{P}$ can be broken by introducing
$\hat{H}_{\delta}(\mathbf{k})$ to Eq. $\text{(\ref{EqCubic}})$,
where $H_{\delta}(\mathbf{k})$ is defined as
\begin{equation}
\hat{H}_{\delta}(\mathbf{k})=\delta\sum_{i}k_{i}\left(\hat{J}_{i+1}\hat{J}_{i}\hat{J}_{i+1}-\hat{J}_{i+2}\hat{J}_{i}\hat{J}_{i+2}\right),\label{eq:Brydon Hamiltonian}
\end{equation}
with $\delta$ denoting the strength of anti-symmetric spin-orbit
coupling (ASOC). This term originates from $T_{d}$ point group symmetry.
Note that both $\hat{C}_{2,x+y}$ and $\hat{P}$ symmetries are absent
in $\hat{H}_{\delta}(\mathbf{k})$ since 
\begin{align}
\hat{C}_{2,x+y}\hat{H}_{\delta}(\mathbf{k})\hat{C}_{2,x+y}^{\dagger} & \neq\hat{H}_{\delta}(k_{y},k_{x},-k_{z}),\\
\hat{P}\hat{H}_{\delta}(\mathbf{k})\hat{P}^{\dagger} & \neq\hat{H}_{\delta}(\mathbf{k}).
\end{align}
Consequently, $\hat{C}_{T}$ is broken due to absence of $\hat{C}_{2,x+y}$
symmetry. In this case, the low-energy superconducting nodes are lifted
along the TPT directions for a sufficiently large $\delta$ (of the
same order of symmetric spin-orbit coupling), and the topological
surface states become unstable. To illustrate this, the excitation
spectra are plotted in Figs. $\text{\ref{Disorders}}$ (c1-c3) in
the presence of $H_{\delta}(\mathbf{k})$. Clearly, the surface states
emerging from the finite energy gap closing point hybridize with the
bulk states at large momentum. This can be understood from Eq. (6)
since then the block digonalization is no applicable.

This analysis identifies $\hat{C}_{T}$ symmetry is required for the
stability of the surface states.

\section{Helical Dirac surface states at finite energies\label{sec:Helical-Dirac-surface}}

In this section, we provide details for the derivation of the helical
Dirac surface states described in the main text. The BdG Hamiltonian
along the {[}001{]} direction becomes
\begin{equation}
\hat{H}(k_{z})=\left(\begin{array}{cccc}
\hat{H}_{1} & 0 & \hat{H}_{3} & 0\\
0 & \hat{H}_{2} & 0 & \hat{H}_{3}\\
\hat{H}_{3} & 0 & -\hat{H}_{1} & 0\\
0 & \hat{H}_{3} & 0 & -\hat{H}_{2}
\end{array}\right),
\end{equation}
with
\begin{equation}
\hat{H}_{1}=\text{diag}(E_{k_{z}}^{+},E_{k_{z}}^{-}),\ \hat{H}_{2}=\text{diag}(E_{k_{z}}^{-},E_{k_{z}}^{+}),
\end{equation}
where $E_{k_{z}}^{+}=(\alpha+9\beta/4)k_{z}^{2}-\mu$, $E_{k_{z}}^{-}=(\alpha+\beta/4)k_{z}^{2}-\mu$,
$\hat{H}_{3}\!\!=\!\!\boldsymbol{\Delta}k_{z}\hat{\sigma}_{x}$ ,
and $\boldsymbol{\Delta}=\sqrt{3}\Delta/2$. To be able to apply our
theory, presented in Eqs. (2-6) of the main text, we should represent
$\hat{H}(k_{z})$ in the pseudospin basis where each diagonal block
contains a pair of doubly degenerate bands. This can be done through
the unitary transformation $\hat{\mathscr{U}}_{1}$
\begin{align}
\!\hat{H}_{1}(k_{z})\! & =\!\hat{\mathscr{U}}_{1}^{\dagger}H(k_{z})\hat{\mathscr{U}}_{1}\!\\
 & =\!\!\left(\!\!\begin{array}{cccc}
E_{k_{z}}^{+} & \!0 & 0 & \!\boldsymbol{\Delta}k_{z}\\
0 & \!E_{k_{z}}^{-} & \boldsymbol{\Delta}k_{z} & \!0\\
0 & \!\boldsymbol{\Delta}k_{z} & -E_{k_{z}}^{+} & \!0\\
\boldsymbol{\Delta}k_{z} & \!0 & 0 & -E_{k_{z}}^{-}
\end{array}\!\!\right)\!\otimes\!\hat{\sigma}_{0},\!\!\!\!\!\!\!\label{PseudoSPin}
\end{align}
where the unitary matrix $\hat{\mathscr{U}}_{1}^{\dagger}=\hat{\mathscr{U}}_{1}^{-1}$
is given by
\begin{equation}
\hat{\mathscr{U}}_{1}=\left(\begin{array}{cccccccc}
1 & 0 & 0 & 0 & 0 & 0 & 0 & 0\\
0 & 0 & 1 & 0 & 0 & 0 & 0 & 0\\
0 & 0 & 0 & 1 & 0 & 0 & 0 & 0\\
0 & 1 & 0 & 0 & 0 & 0 & 0 & 0\\
0 & 0 & 0 & 0 & 1 & 0 & 0 & 0\\
0 & 0 & 0 & 0 & 0 & 0 & 1 & 0\\
0 & 0 & 0 & 0 & 0 & 0 & 0 & 1\\
0 & 0 & 0 & 0 & 0 & 1 & 0 & 0
\end{array}\right).
\end{equation}
In the next step, Eq. ($\text{\ref{PseudoSPin}}$) should be represented
in the interband basis. This can be done through another unitary transformation
$\hat{\mathscr{U}}_{2}$ as
\begin{align}
\hat{\mathscr{H}}(k_{z}) & =\hat{\mathscr{U}}_{2}^{\dagger}\hat{H}_{1}(k_{z})\hat{\mathscr{U}}_{2}=(\hat{\mathscr{U}}_{1}\hat{\mathscr{U}}_{2})^{\dagger}H(k_{z})\hat{\mathscr{U}}_{1}\hat{\mathscr{U}}_{2}\\
 & =\left(\begin{array}{cc}
\hat{\mathsf{H}}_{k_{z}}^{+-} & 0\\
0 & \hat{\mathsf{H}}_{k_{z}}^{-+}
\end{array}\right),\label{FEblocks}
\end{align}
where
\begin{align}
\!\hat{\mathsf{H}}_{k_{z}}^{+-}\! & =\!\left(\!\begin{array}{cc}
E_{k_{z}}^{+} & \boldsymbol{\Delta}k_{z}\\
\boldsymbol{\Delta}k_{z} & -E_{k_{z}}^{-}
\end{array}\!\right)\!\otimes\!\hat{\sigma}_{0},\!\!\label{MasMas1}\\
\!\hat{\mathsf{H}}_{k_{z}}^{-+}\! & =\!\left(\!\begin{array}{cc}
E_{k_{z}}^{-} & \boldsymbol{\Delta}k_{z}\\
\boldsymbol{\Delta}k_{z} & -E_{k_{z}}^{+}
\end{array}\!\right)\!\otimes\!\hat{\sigma}_{0}.\!\!\label{MasMas2}
\end{align}
The transformation $\hat{\mathscr{U}}_{2}$ fulfills the unitary property
$\hat{\mathscr{U}}_{2}^{\dagger}=\hat{\mathscr{U}}_{2}^{-1}$. Its
explicit matrix form is given by
\begin{equation}
\hat{\mathscr{U}}_{2}=\left(\begin{array}{cccccccc}
0 & 1 & 0 & 0 & 0 & 0 & 0 & 0\\
1 & 0 & 0 & 0 & 0 & 0 & 0 & 0\\
0 & 0 & 0 & 0 & 0 & 1 & 0 & 0\\
0 & 0 & 0 & 0 & 1 & 0 & 0 & 0\\
0 & 0 & 0 & 0 & 0 & 0 & 0 & 1\\
0 & 0 & 0 & 0 & 0 & 0 & 1 & 0\\
0 & 0 & 0 & 1 & 0 & 0 & 0 & 0\\
0 & 0 & 1 & 0 & 0 & 0 & 0 & 0
\end{array}\right).
\end{equation}
Note that $\hat{\mathscr{H}}(k_{z})$, given in Eq. ($\text{\ref{FEblocks}}$),
is identical to Eq. (2) of the main text. In this case, $\boldsymbol{\Delta}k_{z}\hat{\sigma}_{0}$
is the FE pairing potential in Eqs. ($\text{\ref{MasMas1}}$) and
Eqs. ($\text{\ref{MasMas2}}$). The zeros on the off-diagonal blocks
of Eq. $\text{(\ref{FEblocks}})$ indicate vanishing intraband pairing
giving rise to the presence of nodes at the Fermi energy. The diagonal
blocks in $\hat{\mathscr{H}}(k_{z})$ exhibit pseudospin-$\pi$ rotation
symmetry along the z(x)-axis, i.e., the symmetry relations given in
Eqs. ($\text{\ref{FE pairing 1}}$) and ($\text{\ref{FE pairing 3}}$).
The explicit matrix form for such an operator can be obtained as
\[
\hat{\mathcal{D}}_{\mathbf{n}_{z}}(\pi,-1)=\left(\begin{array}{cccc}
0 & i & 0 & 0\\
i & 0 & 0 & 0\\
0 & 0 & 0 & i\\
0 & 0 & i & 0
\end{array}\right).
\]
Thus, Eq. $(\text{\ref{FEblocks}})$ is further reducible in the eigenspace
of $\hat{\mathcal{D}}_{\mathbf{n}_{z}}(\pi,-1)$. This can be done
through the unitary matrix $\hat{\mathscr{U}}_{3}$
\begin{align}
\hat{\mathscr{H}}^{\prime}(k_{z}) & =\hat{\mathscr{U}}_{3}^{\dagger}\hat{\mathscr{H}}(k_{z})\hat{\mathscr{U}}_{3}\\
 & =(\hat{\mathscr{U}}_{1}\hat{\mathscr{U}}_{2}\hat{\mathscr{U}}_{3})^{\dagger}H(k_{z})\hat{\mathscr{U}}_{1}\hat{\mathscr{U}}_{2}\hat{\mathscr{U}}_{3}\\
 & =\text{diag}(\hat{h}_{k_{z},+}^{+-},\hat{h}_{k_{z},+}^{+-},\hat{h}_{k_{z},+}^{-+},\hat{h}_{k_{z},-}^{-+}),\label{Diagonal}
\end{align}
with
\begin{equation}
\!\hat{h}_{k_{z},\pm}^{+-}\!=\!\left(\!\begin{array}{cc}
-E_{k_{z}}^{-} & \boldsymbol{\Delta}k_{z}\\
\boldsymbol{\Delta}k_{z} & E_{k_{z}}^{+}
\end{array}\!\right)\!,\ \hat{h}_{k_{z},\pm}^{-+}\!=\!\left(\!\begin{array}{cc}
-E_{k_{z}}^{+} & \boldsymbol{\Delta}k_{z}\\
\boldsymbol{\Delta}k_{z} & E_{k_{z}}^{-}
\end{array}\!\right)\!,\!
\end{equation}
and $\hat{\mathscr{U}}_{3}=\text{diag}(\hat{\mathscr{V}},\hat{\mathscr{V}})$
where $\hat{\mathscr{V}}$ is the matrix of eigenvectors of $\hat{\mathcal{D}}_{\mathbf{n}_{z}}(\pi,-1)$
given by
\[
\hat{\mathscr{V}}=\frac{1}{\sqrt{2}}\left(\begin{array}{cccc}
0 & 1 & 0 & -1\\
0 & 1 & 0 & 1\\
1 & 0 & -1 & 0\\
1 & 0 & 1 & 0
\end{array}\right).
\]
We solve the eigenvalue problem for one of the decoupled subblocks
in Eq. $\text{\eqref{Diagonal}}$. This problem is given by $\hat{h}_{k_{z},\pm}^{-+}\hat{\Phi}(\xi,z)=\mathscr{E}_{\text{DP}}\hat{\Phi}(\xi,z)$
where $\hat{\Phi}(\xi,z)=(u,v)^{T}\text{exp}(\xi z)$ is the ansatz
for the decaying eigenspinor. $\xi$ denotes the localization factor,
$|u|^{2}$ ($|v|^{2}$) is the probability weight for electron (hole)
states with different magnetic quantum number $m_{j}=\pm3/2$ ($m_{j}=\pm1/2$).
We consider a semi-infinite system in $z\geq0$ space. Therefore,
$k_{z}$ is no longer conserved and we use its real space representation
$k_{z}=k_{z}^{\dagger}=-i\partial_{z}$. In this case, the secular
equation $(\hat{h}_{-i\partial_{z},\pm}^{-+}-\mathscr{E}_{\text{DP}})\hat{\Phi}(\xi,z)=0$
can be evaluated by setting its determinant to zero,
\begin{equation}
\Bigg|\begin{array}{cc}
m^{\prime}\xi^{2}+\mu-\mathscr{E}_{\text{DP}} & -i\boldsymbol{\Delta}\xi\\
-i\boldsymbol{\Delta}\xi & -(m\xi^{2}+\mu)-\mathscr{E}_{\text{DP}}
\end{array}\Bigg|=0,\label{DetSec}
\end{equation}
where $m=\alpha+\beta/4$ and $m^{\prime}=\alpha+9\beta/4$.\textcolor{red}{{}
}The solution of the secular equation yields
\begin{align}
\xi_{\pm}= & \sqrt{\frac{1}{2mm^{\prime}}\Big(\Lambda\pm\sqrt{\Lambda^{2}+4mm^{\prime}(\mathscr{E}_{\text{DP}}^{2}-\mu^{2})}\Big)},
\end{align}
where $\Lambda=\boldsymbol{\Delta}^{2}-\mathscr{E}_{\text{DP}}\vartheta^{-}-\mu\vartheta^{+}$
and $\vartheta^{\pm}=(m^{\prime}\pm m)/2$. In the absence of pairing,
i.e., for $\boldsymbol{\Delta}=0$, the localization length becomes
purely imaginary leading to extended states. However, $\xi_{\pm}$
obtains a real component in the topological phase induced by FE pairing.
This leads to proper surface state solutions.

To specify the components of $\hat{\Phi}(\xi,z)$, we use the secular
equation, and obtain
\begin{align}
u\equiv u_{\iota}=-i\boldsymbol{\Delta}\xi_{\iota}\rightarrow v\equiv v_{\iota} & =m^{\prime}\xi_{\iota}^{2}+\mu-\mathscr{E}_{\text{DP}},\\
v\equiv V_{\iota}=-i\boldsymbol{\Delta}\xi_{\iota}\rightarrow u\equiv U_{\iota} & =m\xi_{\iota}^{2}+\mu+\mathscr{E}_{\text{DP}}.
\end{align}
Consequently, we have a pair of eigenspinors given by
\begin{align}
\!\hat{\Phi}(\xi_{\iota},z) & =\Bigg(\begin{array}{c}
u_{\iota}\\
v_{\iota}
\end{array}\Bigg)e^{\xi_{\iota}z},\ \ \hat{\Phi}^{\prime}(\xi_{\iota},z)=\Bigg(\begin{array}{c}
U_{\iota}\\
V_{\iota}
\end{array}\Bigg)e^{\xi_{\iota}z}\!.\label{Ansatz1-1}
\end{align}
We can construct two sets of wave functions by the superposition of
eigenspinors $\hat{\Phi}(\xi_{\iota},z)$ $(\hat{\Phi}^{\prime}(\xi,z))$,
\begin{equation}
\hat{\mathsf{\Psi}}_{1}(z)\!\!=\!\!\sum_{\iota=\pm}C_{\iota}\hat{\Phi}(\xi_{\iota},z),\ \ \ \hat{\mathsf{\Psi}}_{2}(z)\!\!=\!\!\sum_{\iota=\pm}Q_{\iota}\hat{\Phi}^{\prime}(\xi_{\iota},z),\label{GeneralWavefunction}
\end{equation}
where the summations run over the decay factors and the coefficients
of the expansion are denoted by $C_{\iota}$ and $Q_{\iota}$. To
have surface state solutions, the wave functions and their first derivatives
must vanish at the interface of the system and far away from the interface,
i.e.,
\begin{equation}
\hat{\mathsf{\Psi}}(\infty)=\hat{\mathsf{\Psi}}^{\prime}(\infty)=0,\ \ \hat{\mathsf{\Psi}}(0)=\hat{\mathsf{\Psi}}^{\prime}(0)=0.\label{Hardwall1}
\end{equation}
Note that we assume just one interface in our analytical calculations.
The boundary conditions at $z=0$ give rise to two pairs of equations
\begin{align}
C_{-} & \Bigg(\begin{array}{c}
u_{-}\\
v_{-}
\end{array}\Bigg)+C_{+}\Bigg(\begin{array}{c}
u_{+}\\
v_{+}
\end{array}\Bigg)=0,\label{Secular1-1}\\
Q_{-} & \Bigg(\begin{array}{c}
U_{-}\\
V_{-}
\end{array}\Bigg)+Q_{+}\Bigg(\begin{array}{c}
U_{+}\\
V_{+}
\end{array}\Bigg)=0.\label{Secular2-1}
\end{align}
Re-arranging Eqs. $\text{\eqref{Secular1-1}}$ and $\text{\eqref{Secular2-1}}$,
we arrive at
\begin{align}
\frac{C_{-}}{C_{+}} & =-\frac{\xi_{+}}{\xi_{-}}=-\frac{m^{\prime}\xi_{+}^{2}\!+\!\mu\!-\mathscr{E}_{\text{DP}}}{m^{\prime}\xi_{-}^{2}\!+\!\mu\!-\mathscr{E}_{\text{DP}}},\label{Secular3}\\
\frac{Q_{-}}{Q_{+}} & =-\frac{\xi_{+}}{\xi_{-}}=-\frac{m\xi_{+}^{2}\!+\!\mu\!+\mathscr{E}_{\text{DP}}}{m\xi_{-}^{2}\!+\!\mu\!+\mathscr{E}_{\text{DP}}}.\label{Secular4}
\end{align}
Combining Eqs. $\text{\eqref{Secular3}}$ and $\text{\eqref{Secular4}}$
results in the explicit formula for the energy of the helical Dirac
surface points $\mathscr{E}_{\text{DP}}$ given in Eq. (7) of the
main text.

Choosing $C_{-}=Q_{-}=\xi_{+}$ and $C_{+}=Q_{+}=-\xi_{-}$, this
allows us to derive the general eigenfunction corresponding to $\mathscr{E}_{\text{DP}}$,
\begin{equation}
\hat{\mathsf{\Psi}}(z)=\mathcal{C}\xi_{+}\Bigg(\begin{array}{c}
i\boldsymbol{\Delta}\xi_{-}\\
m^{\prime}\xi_{-}^{2}+\mu-\mathscr{E}_{\text{DP}}
\end{array}\Bigg)(e^{-\xi_{-}z}-e^{-\xi_{+}z}),\label{General Wave function2}
\end{equation}
where $\mathcal{C}$ is the normalization factor
\begin{equation}
\mathcal{C}=\frac{1}{\sqrt{\left(|\kappa_{1}|^{2}+|\kappa_{2}|^{2}\right)}}\frac{1}{\sqrt{\int_{0}^{\infty}dz\ |f(z)|^{2}}}.
\end{equation}

\section{Effective 2D helical surface Hamiltonian\label{Sec:EffectHam}}

In this section, we derive the effective Hamiltonian for the 2D helical
surface states given in Eq. (8) of the main text. To do so, we need
to project the bare BdG Hamiltonian onto the helical Dirac surface
states basis. Note that $\hat{\mathsf{\Psi}}(z)$ in Eq. $\text{\eqref{General Wave function2}}$
is the eigenfunction corresponding to the subblock matrix $\hat{h}_{{\bf k},+}^{-+}$,
and we have defined $f(z)\equiv(e^{-\xi_{-}z}-e^{-\xi_{+}z})$, $\kappa_{1}\equiv i\boldsymbol{\Delta}\xi_{+}\xi_{-}$,
and $\kappa_{2}\equiv\xi_{+}(m^{\prime}\xi_{-}^{2}+\mu-\mathscr{E}_{\text{DP}})$.
To have a proper projection basis, we also need the eigenfunction
for the subblock $\hat{h}_{{\bf k},+}^{+-}$. It is given by $\hat{\varphi}(z)=\mathcal{C}(\kappa_{3},\kappa_{4})^{T}\gamma(z)$
where $\kappa_{3}\equiv\xi_{+}(m\xi_{-}^{2}+\mu+\mathscr{E}_{\text{DP}})$,
$\kappa_{4}\equiv-i\varDelta\xi_{+}\xi_{-}$, and $\gamma(z)=-f(z)$.

$\hat{\mathsf{\Psi}}(z)$ and $\hat{\varphi}(z)$ are $2\times1$
column vectors. In order to use them for the projection method, we
convert them to $8\times1$ representation since the BdG Hamiltonian
is a $8\times8$ matrix. This can be done through the transformation
made by the $8\times2$ columns of the matrix $\hat{\mathscr{U}}^{-1}=\{\hat{\gamma}_{1},\hat{\gamma}_{2},\hat{\gamma}_{3},\hat{\gamma}_{4}\}$.
The first and forth subblock matrices in Eq. $\text{\eqref{Diagonal}}$
are identical corresponding to $\hat{h}_{{\bf k},+}^{-+}$ and $\hat{h}_{{\bf k},-}^{-+}$,
respectively. Thus, they correspond to doubly degenerate helical surface
states with eigenvalue $\mathscr{E}_{\text{DP}}$. Therefore, their
$8\times1$ representations take the form
\begin{align}
\hat{\Gamma}_{1}^{-} & \equiv\hat{\gamma}_{1}\hat{\mathsf{\Psi}}(z)=\mathcal{C}f(z)(0,\kappa_{2},0,0,\kappa_{1},0,0,0)^{T},\label{G1}\\
\hat{\Gamma}_{2}^{-} & \equiv\hat{\gamma}_{4}\hat{\mathsf{\Psi}}(z)=\mathcal{C}f(z)(0,0,\kappa_{2},0,0,0,0,\kappa_{1})^{T}.\label{G2}
\end{align}
We repeat the above steps for $\hat{\varphi}(z)$ to obtain the proper
basis for the sectors $\hat{h}_{{\bf k},+}^{+-}$ and $\hat{h}_{{\bf k},-}^{+-}$,
\begin{align}
\hat{\Gamma}_{1}^{+} & \equiv\hat{\gamma}_{2}\hat{\varphi}(z)=\mathcal{C}\gamma(z)(\kappa_{4},0,0,0,0,\kappa_{3},0,0)^{T},\label{G3}\\
\hat{\Gamma}_{2}^{+} & \equiv\hat{\gamma}_{3}\hat{\varphi}(z)=\mathcal{C}\gamma(z)(0,0,0,\kappa_{4},0,0,\kappa_{3},0)^{T}.\label{G4}
\end{align}
We use Eqs. (\ref{G1}-\ref{G4}) as the proper orthonormal set of
eigenfunctions to project the bulk superconducting Hamiltonian to
the surface. Note that the orthonormality condition reads $\int_{0}^{\infty}dz[\hat{\mathscr{Y}}^{\nu}]^{\dagger}\hat{\mathscr{Y}}^{\nu}=\hat{\sigma}_{0}$
where $\hat{\mathscr{Y}}^{\pm}=\{\hat{\Gamma}_{1}^{\pm},\hat{\Gamma}_{2}^{\pm}\}$.

To derive the effective Hamiltonian for the 2D helical surface states,
we consider the conserved wave vectors $k_{x}$ and $k_{y}$ in the
BdG Hamiltonian to be small close to the $\Gamma$ point. Then, we
project $\hat{H}(k_{x},k_{y},-i\partial_{z})$ onto the basis of the
helical Dirac surface states. Thus, the effective Hamiltonian at finite
excitation energies becomes
\begin{table*}
\begin{centering}
\begin{tabular}{|c|c|c|c|c|c|c|}
\hline 
$J$ & $O_{h}$ & $\eta$ & $\hat{\mathcal{O}}_{\eta}(\hat{J})$ & $\hat{\eta}$ & $(L,S)$ & $(L,S)$\tabularnewline
\hline 
$0$ & $A_{1g,u}$ & $A_{1g,u}$ & $\frac{1}{2}\hat{I}_{4}$ & $\hat{A}_{1g,u}\!=\!\hat{\mathcal{N}}_{0,0}$ & $(0,0)$ & $(1,1)$\tabularnewline
\hline 
$1$ & $T_{1u}$ & $T_{1u}^{(1)}$ & $\frac{1}{\sqrt{5}}\hat{J}_{z}$ & $\hat{T}_{1u}^{(1)}\!=\!\hat{\mathcal{N}}_{1,0}$ & $\times$ & $(1,1)$\tabularnewline
 &  & $T_{1u}^{(2)}$ & $\frac{1}{\sqrt{5}}\hat{J}_{y}$ & $\hat{T}_{1u}^{(2)}\!=\!\frac{i}{\sqrt{2}}(\hat{\mathcal{N}}_{1,-1}+\hat{\mathcal{N}}_{1,1})$ & $\times$ & $(1,1)$\tabularnewline
 &  & $T_{1u}^{(3)}$ & $\frac{1}{\sqrt{5}}\hat{J}_{x}$ & $\hat{T}_{1u}^{(3)}\!=\!\frac{1}{\sqrt{2}}(\hat{\mathcal{N}}_{1,-1}-\hat{\mathcal{N}}_{1,1})$ & $\times$ & $(1,1)$\tabularnewline
\hline 
$2$ & $E_{g,u}$ & $E_{g,u}^{(1)}$ & $\frac{1}{6}(3\hat{J}_{z}^{2}-\hat{\mathbf{J}})$ & $\hat{E}_{g,u}^{(1)}\!=\!\hat{\mathcal{N}}_{2,0}$ & $(0,2)$ & $(1,1),(1,3)$\tabularnewline
 &  & $E_{g,u}^{(2)}$ & $\frac{1}{2\sqrt{3}}(\hat{J}_{x}^{2}-\hat{J}_{y}^{2})$ & $\hat{E}_{g,u}^{(2)}\!=\!\frac{1}{\sqrt{2}}(\hat{\mathcal{N}}_{2,-2}+\hat{\mathcal{N}}_{2,2})$ & $(0,2)$ & $(1,1),(1,3)^{*}$\tabularnewline
\hline 
 & $T_{2g,u}$ & $T_{2g,u}^{(1)}$ & $\frac{1}{\sqrt{3}}(\hat{J}_{x}\hat{J}_{y}+\hat{J}_{y}\hat{J}_{x})$ & $\hat{T}_{2g,u}^{(1)}\!=\!\frac{i}{\sqrt{2}}(\hat{\mathcal{N}}_{2,-2}-\hat{\mathcal{N}}_{2,2})$ & $(0,2)$ & $(1,1),(1,3)^{*}$\tabularnewline
 &  & $T_{2g,u}^{(2)}$ & $\frac{1}{\sqrt{3}}(\hat{J}_{z}\hat{J}_{x}+\hat{J}_{x}\hat{J}_{z})$ & $\hat{T}_{2g,u}^{(2)}\!=\!\frac{i}{\sqrt{2}}(\hat{\mathcal{N}}_{2,-1}-\hat{\mathcal{N}}_{2,1})$ & $(0,2)$ & $(1,1),(1,3)^{*}$\tabularnewline
 &  & $T_{2g,u}^{(3)}$ & $\frac{1}{\sqrt{3}}(\hat{J}_{y}\hat{J}_{z}+\hat{J}_{z}\hat{J}_{y})$ & $\hat{T}_{2g,u}^{(3)}\!=\!\frac{i}{\sqrt{2}}(\hat{\mathcal{N}}_{2,-1}+\hat{\mathcal{N}}_{2,1})$ & $(0,2)$ & $(1,1),(1,3)^{*}$\tabularnewline
\hline 
$3$ & $A_{2u}$ & $A_{2u}$ & $\frac{1}{\sqrt{3}}(\hat{J}_{x}\hat{J}_{y}\hat{J}_{z}+\hat{J}_{z}\hat{J}_{y}\hat{J}_{x})$ & $\hat{A}_{2u}\!=\!\frac{i}{\sqrt{2}}(\hat{\mathcal{N}}_{2,-2}-\hat{\mathcal{N}}_{2,2})$ & $\times$ & $(1,3)^{*}$\tabularnewline
\hline 
 & $T_{1u}$ & $T_{1u}^{(1)}$ & $\frac{4}{\sqrt{365}}\hat{J}_{z}^{3}$ & $\hat{T}_{1u}^{(1)}\!=\!\frac{12}{5\sqrt{73}}\hat{\mathcal{N}}_{3,0}$ & $\times$ & $(1,3)$\tabularnewline
 &  & $T_{1u}^{(2)}$ & $\frac{4}{\sqrt{365}}\hat{J}_{y}^{3}$ & $\hat{T}_{1u}^{(2)}\!=\!\frac{3}{5\sqrt{73}i}[\sqrt{5}(\hat{\mathcal{N}}_{3,-3}+\hat{\mathcal{N}}_{3,3})+\sqrt{3}(\hat{\mathcal{N}}_{3,1}+\hat{\mathcal{N}}_{3,-1})]$ & $\times$ & $(1,3)$\tabularnewline
 &  & $T_{1u}^{(3)}$ & $\frac{4}{\sqrt{365}}\hat{J}_{x}^{3}$ & $\hat{T}_{1u}^{(3)}\!=\!\frac{3}{5\sqrt{73}}[\sqrt{5}(\hat{\mathcal{N}}_{3,-3}-\hat{\mathcal{N}}_{3,3})+\sqrt{3}(\hat{\mathcal{N}}_{3,1}-\hat{\mathcal{N}}_{3,-1})]$ & $\times$ & $(1,3)$\tabularnewline
\hline 
 & $T_{2u}$ & $T_{2u}^{(1)}$ & $\frac{1}{\sqrt{3}}\lceil\hat{J}_{z}(\hat{J}_{x}^{2}-\hat{J}_{y}^{2})\rfloor$ & $\hat{T}_{2u}^{(1)}\!=\!\frac{1}{\sqrt{2}}(\hat{\mathcal{N}}_{3,-2}+\hat{\mathcal{N}}_{3,2})$ & $\times$ & $(1,3)^{*}$\tabularnewline
 &  & $T_{2u}^{(2)}$ & $\frac{1}{\sqrt{3}}\lceil\hat{J}_{x}(\hat{J}_{y}^{2}-\hat{J}_{z}^{2})\rfloor$ & $\hat{T}_{2u}^{(2)}\!=\!\frac{1}{4}[\sqrt{3}(\hat{\mathcal{N}}_{3,3}-\hat{\mathcal{N}}_{3,-3})+\sqrt{5}(\hat{\mathcal{N}}_{3,1}-\hat{\mathcal{N}}_{3,-1})]$ & $\times$ & $(1,3)^{*}$\tabularnewline
 &  & $T_{2u}^{(3)}$ & $\frac{1}{\sqrt{3}}\lceil\hat{J}_{y}(\hat{J}_{z}^{2}-\hat{J}_{x}^{2})\rfloor$ & $\hat{T}_{2u}^{(3)}\!=\!\frac{1}{4i}[\sqrt{3}(\hat{\mathcal{N}}_{3,3}+\hat{\mathcal{N}}_{3,-3})-\sqrt{5}(\hat{\mathcal{N}}_{3,1}+\hat{\mathcal{N}}_{3,-1})]$ & $\times$ & $(1,3)^{*}$\tabularnewline
\hline 
\end{tabular}
\par\end{centering}
\caption{\label{TablePairing matrices}Decomposition of total angular momentum
$J$ (first column) in the irreducible representation (irrep) of $O_{h}$
symmetry (second column). The dimension of an irrep is distinguished
by the number of components $\eta$ given in the third column. The
fourth column denotes the normalized irreducible basis matrices for
a component of $O_{h}$ symmetry in $j=3/2$ representation. The fifth
column demonstrates the correspondence between the matrix form for
components of a cubic irrep $\hat{\eta}$ and the components of $SO(3)$
symmetry, namely $\hat{\mathcal{N}}_{J,m_{j}}$ being the total angular
momentum tensor matrices. The last two columns indicate the spin $S$
and orbital $L$ angular momenta of Cooper pairs associated to $J$
and components of a given irrep. The Fermi statistics forces both
$L$ and $S$ to be even ($g$) or odd ($u$). The symmetrization
of the basis matrices is denoted by $\lceil\hat{A}\hat{B}\hat{C}\rfloor=(\hat{A}\hat{B}\hat{C}+\hat{A}\hat{C}\hat{B}+\hat{B}\hat{C}\hat{A}+\hat{B}\hat{A}\hat{C}+\hat{C}\hat{A}\hat{B}+\hat{C}\hat{B}\hat{A})/3!$.
The pairing channels satisfying TPT at FEs are marked by $(*)$.}
\end{table*}
\begin{align}
\hat{\mathtt{H}}(k_{x},k_{y}) & =\int_{0}^{\infty}dz\hat{\mathscr{Y}}^{\nu\dagger}(z)\hat{H}(k_{x},k_{y},-i\partial_{z})\hat{\mathscr{Y}}^{\nu}(z)\nonumber \\
 & =\left(\begin{array}{cc}
A_{1,1} & A_{1,2}\\
A_{2,1} & A_{2,2}
\end{array}\right),\label{Heffective}
\end{align}
where the matrix elements of $\hat{\mathtt{H}}(k_{x},k_{y})$ are
given by
\begin{align}
A_{i,j} & =\int_{0}^{\infty}dz\hat{\Gamma}_{i}^{\dagger}(z)\hat{H}(k_{x},k_{y},-i\partial_{z})\hat{\Gamma}_{j}(z),
\end{align}
with $i,j\in\{1,2\}$. $\hat{\mathtt{H}}(k_{x},k_{y})$ should be
Hermitian, thus, the matrix elements in Eq. $\text{\eqref{Heffective}}$
must fulfill the relations
\begin{align}
A_{1,1} & =A_{1,1}^{*}=A_{2,2}=A_{2,2}^{*},\\
A_{1,2} & =A_{2,1}^{*},\ \ A_{2,1}=A_{1,2}^{*}.
\end{align}
After straightforward algebra, we arrive at the 2D effective Hamiltonian
for the helical surface states given in Eq. (8) of the main text.
Note that the group velocity of the helical topological Dirac surface
states at finite excitation energies takes the form
\begin{equation}
\varsigma_{2}=\frac{\sqrt{3}}{4}\frac{\Delta}{(|\kappa_{1}|^{2}+|\kappa_{2}|^{2})}\text{Im}\left[\left(\kappa_{2}\kappa_{1}^{*}-\kappa_{1}\kappa_{2}^{*}\right)\right].
\end{equation}
It is clear that $\varsigma_{2}$ depends on the components of the
wave function with direct proportionality to the pairing strength
$\Delta$.

\section{Energy scale for finite-energy Cooper pairing in weakly hole-doped
$\text{YPdBi}$\label{Sec:YPdBi}}

\begin{figure}
\centering{}\includegraphics[scale=0.8]{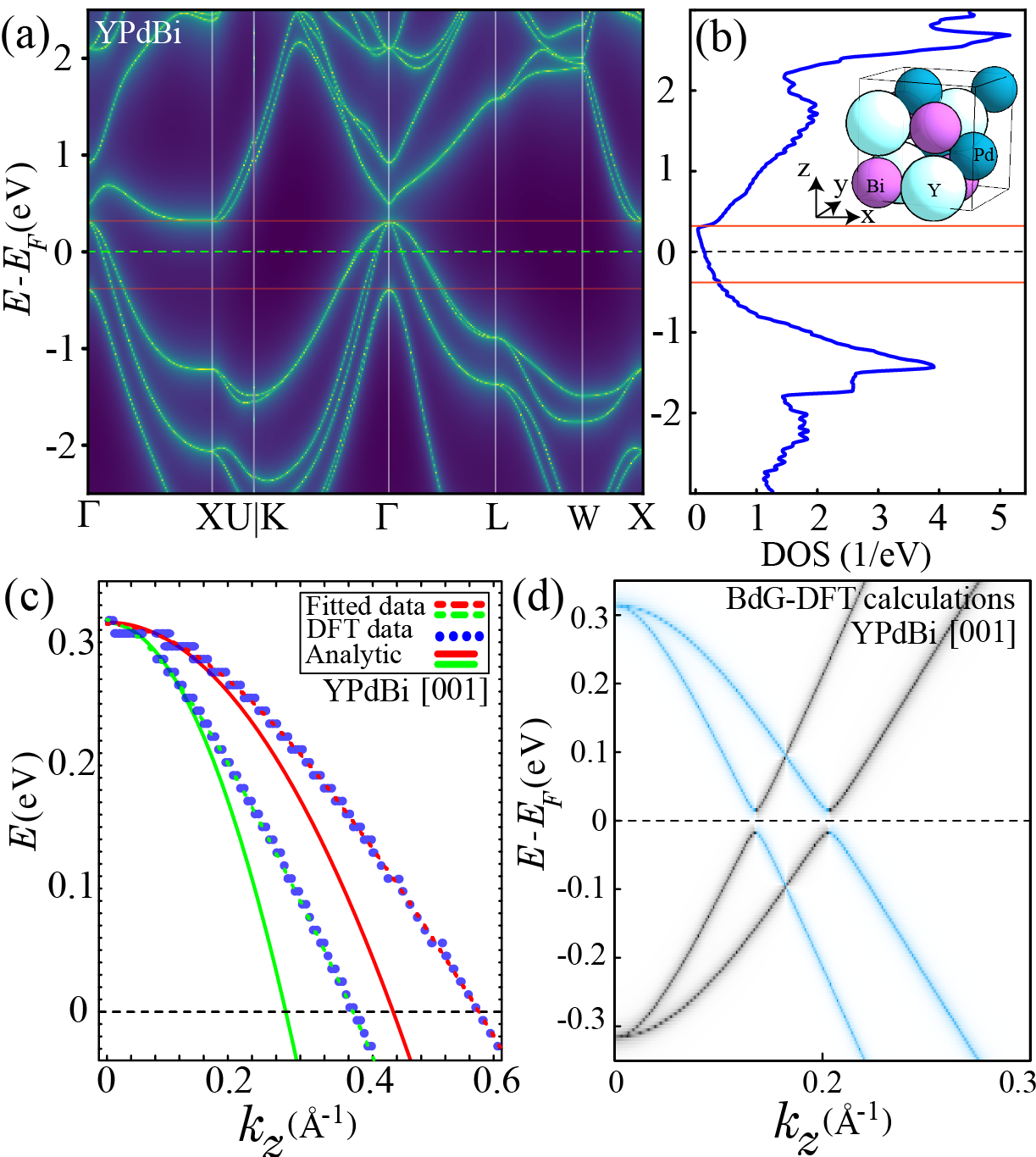}\caption{\label{DFTnormal}(a) DFT normal state band structure for YPdBi. (b)
Density of states for the band structure shown in (a). (c) DFT band
structure in {[}001{]} direction (marked by blue filled circles),
fit to the DFT data (dashed lines) and $\mathbf{k}\cdot\mathbf{p}$
model (solid lines). (d) BdG-DFT band structure in {[}001{]} direction
for $\Delta_{s}\protect\neq0$ and $\Delta_{p}=0$. The blue (black)
coloring of the bands denote their particle (hole) character.}
\end{figure}

In this section, we estimate the energy scale for finite-energy Cooper
pairing in weakly hole-doped YPdBi. We obtain a range of $\Delta_{E}\!\approx\!7.7\!-46.2$~$\textit{\textit{\textmu}}$eV,
which can be resolved in state-of-the-art scanning tunneling microscope
(STM) \citep{LowTemp}. We first refer to the normal state band
structure of YPdBi calculated by density functional theory (DFT),
see Fig.~$\text{\ref{DFTnormal}}$(a).

Close to the $\Gamma$ point, the electronic structure hosts two branches
of the $\Gamma_{8}$ bands curving downward in an energy window around
$\approx0.7$eV before other bands coexists with the $\Gamma_{8}$
bands. This material is a non-centrosymmetric semimetal, which becomes
superconductor at $T_{c}=1.6$ K \citep{Ex2}. The absence
of inversion symmetry results in weak ASOC, which leads to a mixed-parity
superconducting state. Specifically, when the chemical potential resides
in the $j=3/2$ bands, the mixed-parity pairing state $A_{1g}+A_{2u}$
should be favorable in YPdBi due to $T_{d}$ symmetry \citep{TuningParityExp}.

In the following, we estimate the value of finite-energy pairing in
YPdBi. To this end, we employ a combination of DFT and analytical
model analysis \citep{Masoud2023}. Our DFT calculations rely
on the Kohn-Sham-Bogoliubov-de-Gennes (KS-BdG) method as implemented
in the relativistic full-potential JuKKR code \citep{Philipp_1,Philipp_2}.
The crystal structure for YPdBi is taken from the materials project
\citep{Philipp_3_1}, where we use a compressed lattice
constant by 3\% to clearly isolate the $\Gamma_{8}$ bands from the
$\Gamma_{6}$ and $\Gamma_{7}$ bands \citep{DFT_ScientificReports}.
We employ the local density approximation for the normal-state exchange-correlation
functional \citep{Philipp_4} and include the effects of spin-orbit
coupling as well as an angular momentum cutoff of $\ell_{\mathrm{max}}=3$
in the expansion of the basis into spherical harmonics. The Fermi
level is shifted down such that $\mu$ lies in the range where only
the $\Gamma_{8}$ bands persist, as indicated in Fig.~$\text{\ref{DFTnormal}}$(a).
Note that the low density of states in this energy range, see Fig.~$\text{\ref{DFTnormal}}$(b),
allows to tune the chemical potential easily, which could be achievable
experimentally via electron irradiation or by suitable electrical
gating \citep{TuningParityExp}.

To illustrate the need for an unconventional $A_{2u}$ pairing channel
for the existence of finite energy pairing, we first consider only
a constant s-wave pairing channel $A_{1g}$ in our DFT-based KS-BdG
simulations with a (for illustration purposes) large magnitude of
$\Delta_{s}=1$~mRy within the atoms of the YPdBi unit cell.

The normal state and superconducting band structure along the $[001]$
direction are depicted in Figs.~$\text{\ref{DFTnormal}}$(c) and
$\text{\ref{DFTnormal}}$(d), respectively. In both panels, the system
is weakly hole doped, such that the chemical potential lies in the
$j=3/2$ bands, close to the $\Gamma$ point. Clearly, the $j=3/2$
bands have an identical sign of the curvature at the Fermi energy,
i.e., both curl downward. The energy bands are doubly degenerate,
protected by a combination of time-reversal and mirror-reflection
symmetry \citep{CongjunPRB}. Using Eqs.~$(\ref{EqCubic})$ and $(\ref{eq:Brydon Hamiltonian})$,
we obtain the $\mathbf{k}\cdot\mathbf{p}$ spectrum given by
\begin{equation}
E_{\mathbf{k}}^{\pm}=(\alpha+\frac{5}{4}\beta)k_{z}^{2}\pm\beta\sqrt{k_{z}^{4}+\frac{3}{4}\frac{\delta^{2}}{\beta^{2}}k_{z}^{2}}-\mu.\label{YPdBi_energy_bands}
\end{equation}
We fit the DFT data up to the second order polynomials. Then, comparing
these with Eq.~$\text{(\ref{YPdBi_energy_bands}})$, we extract the
model parameters for weakly hole-doped YPdBi close to the $\Gamma$
point as
\begin{align}
\alpha & =-18.1\,\mathring{\text{A}}^{2}\text{eV}, & \beta & =-17.5\,\mathring{\text{A}}^{2}\text{eV},\nonumber \\
\gamma & =16.1\,\mathring{\text{A}}^{2}\text{eV} & \delta & =-0.1\,\mathring{\text{A}}\text{eV},\nonumber \\
\mu & =-317\,\text{meV}.\label{FittedParamters}
\end{align}
Note that the parameter describing ASOC $\delta$ is weak compared
to the symmetric spin-orbit coupling $\beta$, i.e., $\delta/\beta\approx0.00571\,\mathring{\text{A}}^{-1}$.

\begin{figure}
\centering{}\includegraphics[scale=0.42]{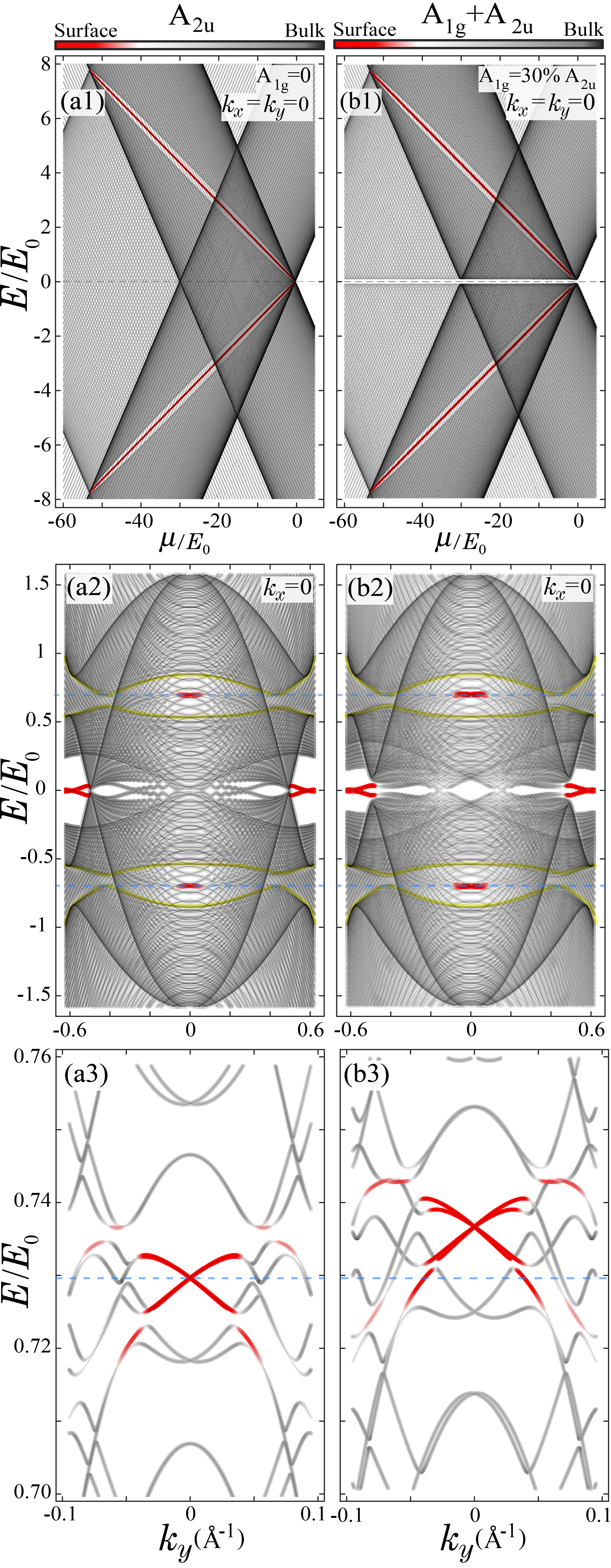}\caption{\label{YPDBIfigu}Excitation spectra for (001) slab of hole-doped
YPdBi with $160$ (layers) obtained by tight-binding regularization
of the effective $\mathbf{k}\cdot\mathbf{p}$ model. The superconducting
state is $A_{1g}+A_{2u}$ where (a1-a3) $\Delta_{s}=0$ and (b1-b3)
$\Delta_{s}=0.3\Delta_{p}$. The surface (bulk) states are marked
by red (gray) colors. The number of layers in panel (a2,b2) {[}(a3,b3){]}
is $200$ {[}$360${]}. The model parameters are the same as those
given in Eq.~$\text{\ensuremath{\left(\ref{FittedParamters}\right)}}$.
The chemical potential is set to $\mu=-5E_{0}$ (hole-doped).
Panels (a3) and (b3) are the enlarged view of the regions marked by
yellow color in panels (a2) and (b2). We have enlarged $\Delta_{p}$
by a factor 3 as compared to normal state energy scales for illustration
reasons.}
\end{figure}

We further analyze the effects of interband pairing in a (001) slab
of YPdBi. The numerical calculations are done by tight-binding regularization
of the $\mathbf{k}\cdot\mathbf{p}$ model given in Eqs.~$\text{(\ref{EqCubic})}$
and $(\ref{eq:Brydon Hamiltonian})$. The results are illustrated
in Fig.~$\ref{YPDBIfigu}$, where we assume open (periodic) boundary
conditions along z-direction (x- and y-directions). In Figs.~$\text{\ref{YPDBIfigu}}$(a1-a3)
and $\text{\ref{YPDBIfigu}}$(b1-b3), the spectra are shown for a
pure odd-parity septet channel $\hat{\Delta}_{\mathbf{k}}^{(A_{2u})}$
and mixed-parity $\Delta_{{\bf k}}=\hat{\Delta}_{{\bf k}}^{(A_{1g})}+\hat{\Delta}_{{\bf k}}^{(A_{2u})}$
channel, respectively. The matrix form for the corresponding pairing
state is explicitly given by

\begin{align}
\hat{\Delta}_{{\bf k}}^{(A_{1g})} & \!=\!\Delta_{s}\!\left(\begin{array}{cccc}
0 & 0 & 0 & 1\\
0 & 0 & -1 & 0\\
0 & 1 & 0 & 0\\
-1 & 0 & 0 & 0
\end{array}\right),\\
\hat{\Delta}_{{\bf k}}^{(A_{2u})} & \!=\!\Delta_{p}\!\left(\!\!\begin{array}{cccc}
\!\frac{3}{4}k_{-} & \!\frac{\sqrt{3}}{2}k_{z} & \!\frac{\sqrt{3}}{4}k_{+} & \!0\\
\!\frac{\sqrt{3}}{2}k_{z} & \!\frac{3}{4}k_{+} & \!0 & \!-\frac{\sqrt{3}}{4}k_{-}\\
\!\frac{\sqrt{3}}{4}k_{+} & \!0 & \!-\frac{3}{4}k_{-} & \!\frac{\sqrt{3}}{2}k_{z}\\
\!0 & \!-\frac{\sqrt{3}}{4}k_{-} & \!\frac{\sqrt{3}}{2}k_{z} & \!-\frac{3}{4}k_{+}
\end{array}\!\!\right)\!,\!\!
\end{align}
where $\Delta_{s(p)}$ denotes the strength for the s(p)-wave $A_{1g}$
($A_{2u}$) pairing channel. Both $\mathbf{k}\cdot\mathbf{p}$ and
DFT calculations imply that interband pairing is absent in the $A_{1g}$
pairing channel, i.e., $\hat{\Delta}_{\mathbf{k}}^{+-}=0$. This is
because the $s$-wave pairing is isotropic in momentum space. However,
interband pairing is present in the $A_{2u}$ channel. This is the
reason for the reduced density of bulk states at finite excitation
energies in the range $|E/E_{0}|\in[0.5,1]$, marked by yellow lines
in Figs.~$\text{\ref{YPDBIfigu}}$(a2) and $\text{\ref{YPDBIfigu}}$(b2).
Using the $\mathbf{k}\cdot\mathbf{p}$ theory, the size of the interband
pairing for the pure $A_{2u}$ channel (in the limit $\delta/\beta\rightarrow0$)
becomes \citep{FinieEnergyPairing-2022-1}
\begin{equation}
\!\!\Delta_{E}\!=\!\sqrt{2}\sqrt{\text{Tr}(\hat{\Delta}_{\tilde{k}_{z}}^{+-}[\hat{\Delta}_{\tilde{k}_{z}}^{+-}]^{\dagger})},\!
\end{equation}
where $\hat{\Delta}_{\tilde{k}_{z}}^{+-}$ denotes the interband pairing
matrix. The relevant electron-hole hybridization takes place at momenta
$\tilde{k}_{z}=\pm2[\mu/(4\alpha+5\beta)]^{1/2}$ and at finite excitation
energies. Along the $[001]$ direction (and equivalent directions),
we obtain $\text{Tr}(\hat{\Delta}_{\tilde{k}_{z}}^{+-}[\hat{\Delta}_{\tilde{k}_{z}}^{+-}]^{\dagger})=3\Delta_{p}^{2}\tilde{k}_{z}^{2}/2$.
In this case, we analytically obtain the size for the gap-like structure
close to the $\Gamma$ point as
\begin{align}
\!\!\Delta_{E} & \!=\!2\Delta_{p}[3\mu/(4\alpha+5\beta)]^{1/2}.\label{EnergyGapSize}
\end{align}
Clearly, $\Delta_{E}$ depends not only on the odd-parity pairing
strength $\Delta_{p}$ but also on the material-dependent model parameters
$\mu,\!\alpha$ and $\beta$.

Moreover, surface states emerge when the chemical potential resides
in the range $\mu/E_{0}\in[-53.46,0]$ where $E_{0}=3\Delta_{p}$,
see Figs.~$\text{\ref{YPDBIfigu}}$(a1) and $\text{\ref{YPDBIfigu}}$(b1).
The surface states are visible as long as the $A_{2u}$ channel dominates
over the $A_{1g}$ channel. Note that we also observe
the superconducting energy gap at the Fermi energy in Fig.~$\text{\ref{YPDBIfigu}}$(b1).
It originates from the isotropic intraband pairing in the $A_{1g}$
channel.

The effects of interband pairing in other directions, where momentum
is conserved, are depicted in Figs.~$\text{\ref{YPDBIfigu}}$(a2)
and $\text{\ref{YPDBIfigu}}$(b2). For the pure $A_{2u}$ state ($\Delta_{s}=0$
and $\Delta_{p}\neq0$), the size for the interband pairing becomes
maximal at the $\Gamma$ point and is analytically given by Eq.~$(\text{\ref{EnergyGapSize}})$.
It decreases monotonically at larger momenta. For the mixed-parity
paring state $A_{1g}+A_{2u}$, the surface states are slightly shifted
upward due to intraband pairing in the $A_{1g}$ channel, compare
Figs.~$\text{\ref{YPDBIfigu}}$(a3) and $\text{\ref{YPDBIfigu}}$(b3).
They also exhibit a weak hybridization with the bulk states at large
momenta, depicted in Figs.~$\text{\ref{YPDBIfigu}}$(b2) and $\text{\ref{YPDBIfigu}}$(b3).

To estimate the magnitude of $\Delta_{p}$ in weakly hole-doped YPdBi,
we assume that it is of similar magnitude as for LuPdBi, where it
is of the order $\Delta_{p}\!=\!50\!-\!300$~$\mathring{\text{A}}\textit{\textit{\textmu}}$eV
\citep{TuningParityExp}. This is because of similarities in the
normal-state band structure, point group symmetries, and the type
of superconducting pairing. Under this assumption, the energy gap
size for the interband pairing becomes $\Delta_{E}\!\approx\!7.7\!-\!46.2$~$\textit{\textit{\textmu}}$eV.
Notably, an energy resolution below 8 $\textit{\textit{\textmu}}$eV
at operating temperatures of 10 mK is achievable in state-of-the-art
STM and transport experiments in dilution refrigerators \citep{LowTemp}.

\section{Pairing matrices}

In this section, we demonstrate the method for obtaining the pairing
matrices in cubic symmetry, given in Table I of the main paper. The
cubic pairing matrices with dependency of momenta up to the linear
order can be derived through the relation between SO(3) \citep{PairingMatrices-2016,PairingMatrices-2015,LucilePRB}
and cubic \citep{LucilePRB,VenderbosPRX,Jiabin,PairingMatrices-2018}
symmetries as we elaborate in the following. In addition, in Ref.
\citep{Multiband-2021-1}, a general group-theoretical method to derive
the symmetry-allowed pairing matrices for any local degrees of freedom
(spin, orbital, and basis site) is introduced. The method relies on
the reduction of product representation of the crystallographic point
group.

The irreducible representations of SO(3) symmetry are labeled by the
total angular momentum $J$, which combines orbital $L$ and spin
$S$ angular momenta. The Cooper pairs formed by two electrons with
$j=3/2$ total angular momentum can have singlet ($J=0$), triplet
($J=1$), quintet ($J=2$), or septet $(J=3)$ angular momenta. Considering
orbital angular momentum up to the p-wave channel, i.e., $L\in\{0,1\}$,
the intrinsic spin angular momentum can take values $S\in\{0,1,2,3\}$
according to relation $|L-S|\leq J\leq|L+S|$. The correspondence
between the components of a cubic irrep and the SO(3) symmetry is
given by the relation \citep{PairingMatrices-2020},
\begin{equation}
\!\!\hat{\eta}=\!\!\!\!\sum_{m_{j_{1}},m_{j_{2}}}\!\sum_{m_{J}}\!\big[\big(\hat{\mathcal{O}}_{\eta}(\hat{J})\hat{\mathcal{R}}\big)_{m_{j_{1}},m_{j_{2}}}\!\langle J,m_{J}|m_{j_{1}},m_{j_{2}}\rangle\big]\hat{\mathcal{N}}_{J,m_{J}},\!\!\label{Correspondece}
\end{equation}
where $\langle J,m_{J}|m_{j_{1}},m_{j_{2}}\rangle$ are the Clebsch-Gordan
coefficients, $m_{j_{1}},m_{j_{2}}\in\{\pm3/2,\pm1/2\}$, $\hat{\mathcal{R}}=e^{i\pi\hat{J}_{y}}$,
$\hat{\mathcal{N}}_{J,m_{J}}$ denotes the multipole matrix for the
total angular momentum labeled with magnetic quantum number $m_{J}\in\{-J,\ldots,J\}$,
and $\hat{\eta}$ is the matrix representation for the component of
an irrep with the normalized basis matrix $\hat{\mathcal{O}}_{\eta}(\hat{J})$
given in the third and fourth columns of Table $\text{\ref{TablePairing matrices}}$,
respectively. Note that the fifth column of Table $\text{\ref{TablePairing matrices}}$
is obtained according to Eq. $\text{\eqref{Correspondece}}$. The
explicit matrix form of $\hat{\mathcal{N}}_{J,m}$ can be derived
by the expansion
\begin{equation}
\hat{\mathcal{N}}_{J,m}=\sum_{m_{L},m_{S}}\langle m_{L},m_{S}|J,m_{J}\rangle Y_{m_{L}}^{L}(\mathbf{k})\hat{\mathcal{S}}_{m_{S}}^{S},\label{Expansion J}
\end{equation}
where $\langle m_{L},m_{S}|J,m_{J}\rangle$ denotes the Clebsch-Gordan
coefficient, $Y_{m_{L}}^{L}(\mathbf{k})$ and $\hat{\mathcal{S}}_{m_{S}}^{S}$
are spherical harmonics and irreducible spin tensor matrices labeled
with $m_{L}\in[-L,\ldots,L]$ and $m_{S}\in[-S,\ldots,S]$ as the
orbital axial angular momentum and spin magnetic quantum number, respectively.
Since we are interested in the odd-parity pairings up to linear momenta,
the orbital angular momentum is $L=1$ ($p$-wave) with the relative
spherical harmonics $Y_{1,\pm1}(\mathbf{k})=\mp\sqrt{3/8\pi}k_{\pm}/|{\bf k}|$
and $Y_{1,0}(\mathbf{k})=\sqrt{3/4\pi}k_{z}/|{\bf k}|$. Moreover,
the derivation method for obtaining $\hat{\mathcal{S}}_{m_{S}}^{S}$
follows Refs. \citep{VenderbosPRX,Jiabin,FinieEnergyPairing-2022-1}.
Using the fifth columns of Table $\text{\ref{TablePairing matrices}}$
and Eq. $\text{\eqref{Expansion J}}$, the cubic pairing matrices
can be derived straightforwardly, giving
\begin{equation}
\hat{\Delta}_{\mathbf{k}}=|\mathbf{k}|\,\hat{\eta}\,\hat{\mathcal{R}}.
\end{equation}

We start with the $A_{2u}$ irrep to find its pairing matrix. $A_{2u}$
is a one-dimensional irrep appearing only for the $J=3$ irrep corresponding
to $L=1$ and $S=3$. The matrix representation for $A_{2u}$ is given
by
\begin{align}
\hat{A}_{2u}\! & =\!\frac{i}{\sqrt{2}}(\hat{\mathcal{N}}_{2,-2}-\hat{\mathcal{N}}_{2,2})\!=\!\frac{\sqrt{2}}{12i}[Y_{1}^{1}(3\hat{\mathcal{S}}_{-3}^{3}-\sqrt{15}\hat{\mathcal{S}}_{1}^{3})\nonumber \\
+ & \sqrt{12}Y_{0}^{1}(\hat{\mathcal{S}}_{-2}^{3}+\hat{\mathcal{S}}_{2}^{3})\!+\!Y_{-1}^{1}(3\hat{\mathcal{S}}_{3}^{3}-\sqrt{15}\hat{\mathcal{S}}_{-1}^{3})],\!\!\!
\end{align}
where we dropped the momentum dependency of spherical harmonics. Spin
multipole matrices, which fulfill the relation $\hat{\mathcal{S}}_{-m_{S}}^{S}=(-1)^{m_{S}}(\hat{\mathcal{S}}_{m_{S}}^{S})^{T}$,
are given by
\begin{equation}
\!\!\hat{\mathcal{S}}_{0}^{3}\!=\frac{\sqrt{5}}{10}\left(\!\!\begin{array}{cccc}
1 & \!\!0 & 0 & 0\\
0 & \!\!-3 & 0 & 0\\
0 & \!\!0 & 3 & 0\\
0 & \!\!0 & 0 & -1
\end{array}\!\!\right)\!,\hat{\mathcal{S}}_{1}^{3}\!=\!\frac{\sqrt{5}}{5}\!\left(\!\!\begin{array}{cccc}
0 & -1 & 0 & \!\!0\\
0 & 0 & \sqrt{3} & \!\!0\\
0 & 0 & 0 & \!\!-1\\
0 & 0 & 0 & \!\!0
\end{array}\!\!\!\right)\!\!,\!\!\!
\end{equation}
and
\begin{equation}
\hat{\mathcal{S}}_{2}^{3}=\frac{\sqrt{2}}{2}\left(\begin{array}{cccc}
0 & 0 & 1 & 0\\
0 & 0 & 0 & -1\\
0 & 0 & 0 & 0\\
0 & 0 & 0 & 0
\end{array}\right)\!,\ \hat{\mathcal{S}}_{3}^{3}=\left(\begin{array}{cccc}
0 & 0 & 0 & -1\\
0 & 0 & 0 & 0\\
0 & 0 & 0 & 0\\
0 & 0 & 0 & 0
\end{array}\right).
\end{equation}
Eventually, the pairing matrix for the $A_{2u}$ irrep takes the form
\citep{Brydon}
\begin{align}
\hat{\Delta}_{{\bf k}} & =|\mathbf{k}|\,\hat{A}_{2u}\,\hat{\mathcal{R}}\nonumber \\
 & =\Delta\left(\begin{array}{cccc}
\frac{3}{4}k_{-} & \frac{\sqrt{3}}{2}k_{z} & \frac{\sqrt{3}}{4}k_{+} & 0\\
\frac{\sqrt{3}}{2}k_{z} & \frac{3}{4}k_{+} & 0 & -\frac{\sqrt{3}}{4}k_{-}\\
\frac{\sqrt{3}}{4}k_{+} & 0 & -\frac{3}{4}k_{-} & \frac{\sqrt{3}}{2}k_{z}\\
0 & -\frac{\sqrt{3}}{4}k_{-} & \frac{\sqrt{3}}{2}k_{z} & -\frac{3}{4}k_{+}
\end{array}\right).
\end{align}
In the next step, we consider the three-dimensional $T_{2u}$ irrep.
It appears for the $J=3$ and $J=5$ irreps. Due the the nature of
$j=3/2$ electron, only the septet component is allowed. In this case,
the first component of the $T_{2u}$ irrep up to linear order of momenta
takes the form
\begin{align}
\hat{T}_{2u}^{(1)}\! & =\!\frac{1}{\sqrt{2}}(\hat{\mathcal{N}}_{3,-2}+\hat{\mathcal{N}}_{3,2})\!=\!\frac{1}{6\sqrt{2}}[Y_{-1}^{1}(\sqrt{15}\hat{\mathcal{S}}_{-1}^{3}+3\hat{\mathcal{S}}_{3}^{3})\nonumber \\
+ & \sqrt{12}Y_{0}^{1}(\hat{\mathcal{S}}_{2}^{3}-\hat{\mathcal{S}}_{-2}^{3})-Y_{1}^{1}(3\hat{\mathcal{S}}_{-3}^{3}+\sqrt{15}\hat{\mathcal{S}}_{1}^{3})].\label{T2u1}
\end{align}
The second component is
\begin{align}
\hat{T}_{2u}^{(2)}= & \frac{1}{4}[\sqrt{3}(\hat{\mathcal{N}}_{3,3}-\hat{\mathcal{N}}_{3,-3})+\sqrt{5}(\hat{\mathcal{N}}_{3,1}-\hat{\mathcal{N}}_{3,-1})]\nonumber \\
= & \frac{1}{24}[\sqrt{3}Y_{-1}^{1}((5\hat{\mathcal{S}}_{2}^{3}-3\hat{\mathcal{S}}_{-2}^{3})-\sqrt{30}\hat{\mathcal{S}}_{0}^{3})\nonumber \\
 & +\sqrt{3}Y_{1}^{1}((5\hat{\mathcal{S}}_{-2}^{3}-3\hat{\mathcal{S}}_{2}^{3})-\sqrt{30}\hat{\mathcal{S}}_{0}^{3})\nonumber \\
 & +Y_{0}^{1}(9(\hat{\mathcal{S}}_{-3}^{3}+\hat{\mathcal{S}}_{3}^{3})+\sqrt{15}(\hat{\mathcal{S}}_{-1}^{3}+\hat{\mathcal{S}}_{1}^{3}))],\label{T2u2}
\end{align}
and the third component
\begin{align}
\hat{T}_{2u}^{(3)}= & \frac{1}{4i}[\sqrt{3}(\hat{\mathcal{N}}_{3,3}+\hat{\mathcal{N}}_{3,-3})-\sqrt{5}(\hat{\mathcal{N}}_{3,1}+\hat{\mathcal{N}}_{3,-1})]\nonumber \\
= & \frac{i}{24}(\sqrt{3}Y_{1}^{1}(3\hat{\mathcal{S}}_{2}^{3}-5\hat{\mathcal{S}}_{-2}^{3}-\sqrt{30}\hat{\mathcal{S}}_{0}^{3})\nonumber \\
 & +\sqrt{3}Y_{-1}^{1}(5\hat{\mathcal{S}}_{2}^{3}-3\hat{\mathcal{S}}_{-2}^{3}+\sqrt{30}\hat{\mathcal{S}}_{0}^{3})\nonumber \\
 & +Y_{0}^{1}(9(\hat{\mathcal{S}}_{-3}^{3}-\hat{\mathcal{S}}_{3}^{3})+\sqrt{15}(\hat{\mathcal{S}}_{1}^{3}-\hat{\mathcal{S}}_{-1}^{3}))).\label{T2u3}
\end{align}
Finally, the explicit pairing matrices for the $T_{2u}$ irrep become
\begin{align}
\hat{\Delta}_{{\bf k}} & =|\mathbf{k}|\,\hat{T}_{2u}^{(1)}\,\hat{\mathcal{R}}\nonumber \\
 & =\!\Delta\!\left(\!\!\begin{array}{cccc}
\frac{3}{4}k_{-} & \frac{\sqrt{3}}{2}k_{z} & \frac{\sqrt{3}}{4}k_{+} & 0\\
\frac{\sqrt{3}}{2}k_{z} & \frac{3}{4}k_{+} & 0 & \frac{1}{4}\sqrt{3}k_{-}\\
\frac{\sqrt{3}}{4}k_{+} & 0 & \frac{3}{4}k_{-} & -\frac{\sqrt{3}}{2}k_{z}\\
0 & \frac{\sqrt{3}}{4}k_{-} & -\frac{\sqrt{3}}{2}k_{z} & \frac{3}{4}k_{+}
\end{array}\!\!\right),\!\!
\end{align}
\begin{align}
\hat{\Delta}_{{\bf k}} & =|\mathbf{k}|\,\hat{T}_{2u}^{(2)}\,\hat{\mathcal{R}}\nonumber \\
 & =\Delta\left(\begin{array}{cccc}
3k_{z} & \frac{1}{\sqrt{3}}k_{-}^{\prime} & \frac{1}{\sqrt{3}}k_{z} & ik_{y}\\
\frac{1}{\sqrt{3}}k_{+}^{\prime} & k_{z} & 3ik_{y} & \frac{1}{\sqrt{3}}k_{z}\\
\frac{1}{\sqrt{3}}k_{z} & 3ik_{y} & k_{z} & \frac{-1}{\sqrt{3}}k_{+}^{\prime}\\
ik_{y} & \frac{1}{\sqrt{3}}k_{z} & \frac{-1}{\sqrt{3}}k_{+}^{\prime} & 3k_{z}
\end{array}\right),
\end{align}
\begin{align}
\hat{\Delta}_{{\bf k}} & =|\mathbf{k}|\,\hat{T}_{2u}^{(3)}\,\hat{\mathcal{R}}\nonumber \\
 & =\Delta\left(\begin{array}{cccc}
3k_{z} & \frac{-1}{\sqrt{3}}k_{-}^{\prime\prime} & \frac{-1}{\sqrt{3}}k_{z} & -k_{x}\\
\frac{-1}{\sqrt{3}}k_{-}^{\prime\prime} & -k_{z} & -3k_{x} & \frac{1}{\sqrt{3}}k_{z}\\
\frac{-1}{\sqrt{3}}k_{z} & -3k_{x} & k_{z} & \frac{-1}{\sqrt{3}}k_{+}^{\prime\prime}\\
-k_{x} & \frac{1}{\sqrt{3}}k_{z} & \frac{-1}{\sqrt{3}}k_{+}^{\prime\prime} & -3k_{z}
\end{array}\right),
\end{align}
where $k_{\pm}^{\prime}=4k_{x}\pm ik_{y}$ and $k_{\pm}^{\prime\prime}=k_{x}\pm4ik_{y}$.

$E_{u}$ $(T_{2u})$ is a two- (three-) dimensional irrep. It appears
for the $J=2$ and $J=4$ irreps. In the $j=3/2$ representation,
only the $J=2$ channel is allowed. Note that the second component
of $E_{u}$ and all the components of $T_{2u}$ can undergo TPTs at
FEs. Thus, the expansion for $\hat{E}_{u}^{(2)}$ is
\begin{align}
\!\!\hat{E}_{u}^{(2)} & \!=\frac{1}{\sqrt{2}}(\hat{\mathcal{N}}_{2,-2}+\hat{\mathcal{N}}_{2,2})\!=\frac{\sqrt{14}}{42}[\sqrt{3}Y_{1}^{1}(\sqrt{15}\hat{\mathcal{S}}_{-3}^{3}+\hat{\mathcal{S}}_{1}^{3})\nonumber \\
+\sqrt{3} & Y_{-1}^{1}(\hat{\mathcal{S}}_{-1}^{3}+\sqrt{15}\hat{\mathcal{S}}_{3}^{3})-\sqrt{15}Y_{0}^{1}(\hat{\mathcal{S}}_{-2}^{3}+\hat{\mathcal{S}}_{2}^{3})].
\end{align}
The pairing matrix for $E_{u}^{(2)}$ becomes
\begin{align}
\hat{\Delta}_{{\bf k}} & =|\mathbf{k}|\,\hat{E}_{u}^{(2)}\,\hat{\mathcal{R}}\nonumber \\
 & =\Delta\left(\begin{array}{c@{\hskip-1pt}c@{\hskip-1pt}ccccccccccccccccccccccccccccccccccccccccccccccccccccccccccccccccccccccccccccccccccccccccccccccccccccccccccccccccccccccccccccccccccccccccccccccccccccccccccccccccccccccccccccccccccccccc}
5\sqrt{3}k_{-} & -5k_{z} & -k_{+} & 0\\
-5k_{z} & -\sqrt{3}k_{+} & 0 & k_{-}\\
-k_{+} & 0 & \sqrt{3}k_{-} & -5k_{z}\\
0 & k_{-} & -5k_{z} & -5\sqrt{3}k_{+}
\end{array}\right).
\end{align}
Likewise, the pairing matrix for the first component of the $T_{2u}$
irrep can be straightforwardly derived
\begin{align}
\!\!\hat{\Delta}_{{\bf k}} & =|\mathbf{k}|\,\hat{T}_{2u}^{(1)}\,\hat{\mathcal{R}}\nonumber \\
 & =\Delta\!\left(\!\!\begin{array}{cccc}
-5\sqrt{3}k_{-} & \!\!5k_{z} & \!\!k_{+} & \!\!0\\
5k_{z} & \!\!\sqrt{3}k_{+} & \!\!0 & \!\!k_{-}\\
k_{+} & \!\!0 & \!\!\sqrt{3}k_{-} & \!\!-5k_{z}\\
0 & \!\!k_{-} & \!\!-5k_{z} & \!\!-5\sqrt{3}k_{+}
\end{array}\!\!\right)\!,\!\!\!\!
\end{align}
and the second component takes the form
\begin{align}
\!\!\hat{\Delta}_{{\bf k}} & =|\mathbf{k}|\,\hat{T}_{2u}^{(2)}\,\hat{\mathcal{R}}\nonumber \\
 & =\Delta\left(\!\!\begin{array}{cccc}
0 & -5k_{-} & 4k_{z} & \sqrt{3}k_{x}\\
-5k_{-} & 4\sqrt{3}k_{z} & 3\sqrt{3}k_{x} & -4k_{z}\\
4k_{z} & 3\sqrt{3}k_{x} & -4\sqrt{3}k_{z} & -5k_{+}\\
\sqrt{3}k_{x} & -4k_{z} & -5k_{+} & 0
\end{array}\!\!\right)\!.\!\!\!
\end{align}
Finally, the pairing potential for the third component of $T_{2u}$
becomes
\begin{align}
\!\!\hat{\Delta}_{{\bf k}} & =|\mathbf{k}|\,\hat{T}_{2u}^{(3)}\,\hat{\mathcal{R}}\nonumber \\
 & =\Delta\!\left(\!\!\begin{array}{cccc}
0 & \!\!-5k_{-} & 4k_{z} & \!\!i\sqrt{3}k_{y}\\
-5k_{-} & \!\!4\sqrt{3}k_{z} & 3i\sqrt{3}k_{y} & \!\!4k_{z}\\
4k_{z} & \!\!3i\sqrt{3}k_{y} & 4\sqrt{3}k_{z} & \!\!5k_{+}\\
i\sqrt{3}k_{y} & \!\!4k_{z} & 5k_{+} & \!\!0
\end{array}\!\!\right).\!\!\!\!
\end{align}

To sum up, the relation between cubic point group symmetry and full
SO(3) symmetry enabled us to explicitly obtain the cubic pairing matrices.
This method can also be applied to other point group symmetries.

\clearpage

\begin{thebibliography}{129}
\bibitem [{Com({\natexlab{c}})}]{Comment_Tightbinding}%
  \BibitemOpen
  \href@noop {} {\  }\bibinfo {note} {\!\!To evaluate the
  tight-binding calculations, we used standard lattice regularizations
  $k_{z}^{2}\rightarrow2(1-\text{cos}(k_{z}a))/a^2$ and
  $k_{z}\rightarrow\text{sin}(k_{z}a)/a$ where $a$ is the lattice
  constant.}\BibitemShut {Stop}%
\bibitem [{Com({\natexlab{b}})}]{Comment_IPR}%
  \BibitemOpen
  \href@noop {} {\ }\bibinfo {note} {\!\!The IPR is defined
  by $I_{E}=\text{ln}(\sum_{i=1}^{8N}|\langle
  i|\psi\rangle|^{4}/\text{ln}(8N))$ where $N$ is the total number of sites and
  $|\psi\rangle$ indicates the eigenvector associated to $E$. The localized
  (extended) boundary (bulk) states have $I_{E}\rightarrow 0\
  (-1)$.}\BibitemShut {Stop}%
\bibitem [{Com({\natexlab{f}})}]{Comment_Hamil}%
  \BibitemOpen
  \href@noop {} {\  }\bibinfo {note} {\!\!Note that $\hat{h}_{{\bf
  k},\lambda}^{-+}$ can be easily derived by substitution of
  $+\leftrightarrow-$ in Eq.\ $\text{\eqref{FinalDecopled}}$. Note that
  $\delta_{{\bf k},\lambda}^{+-}$ fulfills the symmetry relation
  $\delta_{{\bf k},\lambda}^{-+}=-\delta_{{\bf
  -k},\lambda}^{+-}$.}\BibitemShut {Stop}%
\bibitem [{Com({\natexlab{h}})}]{Comment_Table}%
  \BibitemOpen
  \href@noop {} {\ }\bibinfo {note} {\!\!\!The FE topological
  Dirac surface states are not restricted to the $A_{2u}$ channel. Rather, they
  also exist for the other pairing channels summarized in Table
  $\text{\ref{Table1}}$. The TPTs at FEs happen along the $C_{2}$ axes for the
  $E_{u}$ and $T_{2u}$ channels, and along the $C_{2}^{\prime}$ axes for the
  $T_{2u}$ channel.}\BibitemShut {Stop}%
\bibitem [{Com({\natexlab{i}})}]{Comment_FourfoldDeg}%
  \BibitemOpen
  \href@noop {} {\  }\bibinfo {note} {\!\!\!The Dirac points are two-fold degenerate at one surface at positive and at negative excitation energies. Considering two surfaces, we obtain the same set of Dirac points on each surface. Thus, the degeneracy is doubled as compared to the case of single surface.}\BibitemShut {Stop}%
\bibitem [{Com({\natexlab{i}})}]{com_generic}%
  \BibitemOpen
  \href@noop {} {\  }\bibinfo {note} {\!\!\!Although, spin-orbit coupling is an essential
ingredient for the formation of such surface states in our model, the effective masses in systems with nearly vanishing spin-orbit coupling
can stem from different orbitals giving rise to non-vanishing values for $m-m^{\prime}$.}\BibitemShut {Stop}%
\bibitem [{Com({\natexlab{a}})}]{Comment_ASOC}%
  \BibitemOpen
  \href@noop {} {\ }\bibinfo {note}{\!\!\!\!\!The inversion
  breaking anti-symmetric spin-orbit-coupling (ASOC) in half-$\text{H}$eusler
  material is weak compared to symmetric spin-orbit coupling.}\BibitemShut
  {Stop}%
\end{thebibliography}
\end{document}